\documentclass[twocolumn,prl,superscriptaddress]{revtex4}
\usepackage[utf8]{inputenc}
\usepackage{amsmath}
\usepackage{amssymb}
\usepackage{bm}
\usepackage{graphicx}
\usepackage{color}

\begin{document}

\title{Quasiparticle poisoning in trivial and topological Josephson junctions}
\author{Aleksandr E. Svetogorov}
\author{Daniel Loss}
\author{Jelena Klinovaja}
\affiliation{Department of Physics, University of Basel, Klingelbergstrasse, 4056 Basel, Switzerland}

\begin{abstract}
We study theoretically a short single-channel Josephson junction between superconductors in the trivial and topological phases.
The junction is assumed to be biased by a small current and subjected to quasiparticle poisoning. We find that the presence of quasiparticles leads to a voltage signal from the Josephson junction that can be observed both in the trivial and in the topological phase. Quite remarkably, these voltage signatures are sufficiently different in the two phases such that they can serve as means to clearly distinguish between trivial Andreev and topological Majorana bound states in the system. Moreover, these voltage signatures, in the trivial and topological phase, would allow one to measure directly the quasiparticle poisoning rates and to test various approaches for protection against quasiparticle poisoning. 
\end{abstract}
\maketitle

{\it Introduction.} 
Majorana fermions in condensed matter physics have been a `hot' topic in the community ever since the first claim that these non-abelian anyons can exist in mesoscopic systems~\cite{Kitaev2001}. Nevertheless, experimental realization of a system supporting Majorana bound states (MBSs) as well as direct observation of these bound states presents a sophisticated problem. Despite numerous claims of indirect observation of MBSs, there have always been concerns that the demonstrated effects may have different origins. One of the biggest problems is quasiparticle poisoning (QP), which causes a change of the fermion parity, while the  Majorana qubit is based on a fixed-parity model, as well as the majority of effects proposed to establish the presence of MBSs in a system, such as fractional Josephson effect~\cite{Kitaev2001,Kwon2003,Kane2009,Wiedenmann2016}. Despite being studied over the last decade~\cite{Visser2011,Chamon2011,Rainis2012,Trauzettel2012,Freedman2017,Nayak2018,Pikulin2021}, QP still requires both theoretical and experimental attention, as different theoretical approaches lead to different estimates on the QP rates mostly due to accounting for different sources of quasiparticles. In this letter we propose that the high sensitivity of devices with MBSs can actually be used to indicate the topological phase as well as to measure the QP rates and to get a better understanding of the possible mechanisms leading to QP. Moreover, we discuss the effect of poisoning in case of trivial (single-channel) Josephson junctions and show that it is principally different from the effect in the topological phase, which provides an unambiguous way to distinguish Andreev bound states (ABSs) in the trivial phase from MBSs in the topological phase.

\begin{figure}
\includegraphics[width=7cm]{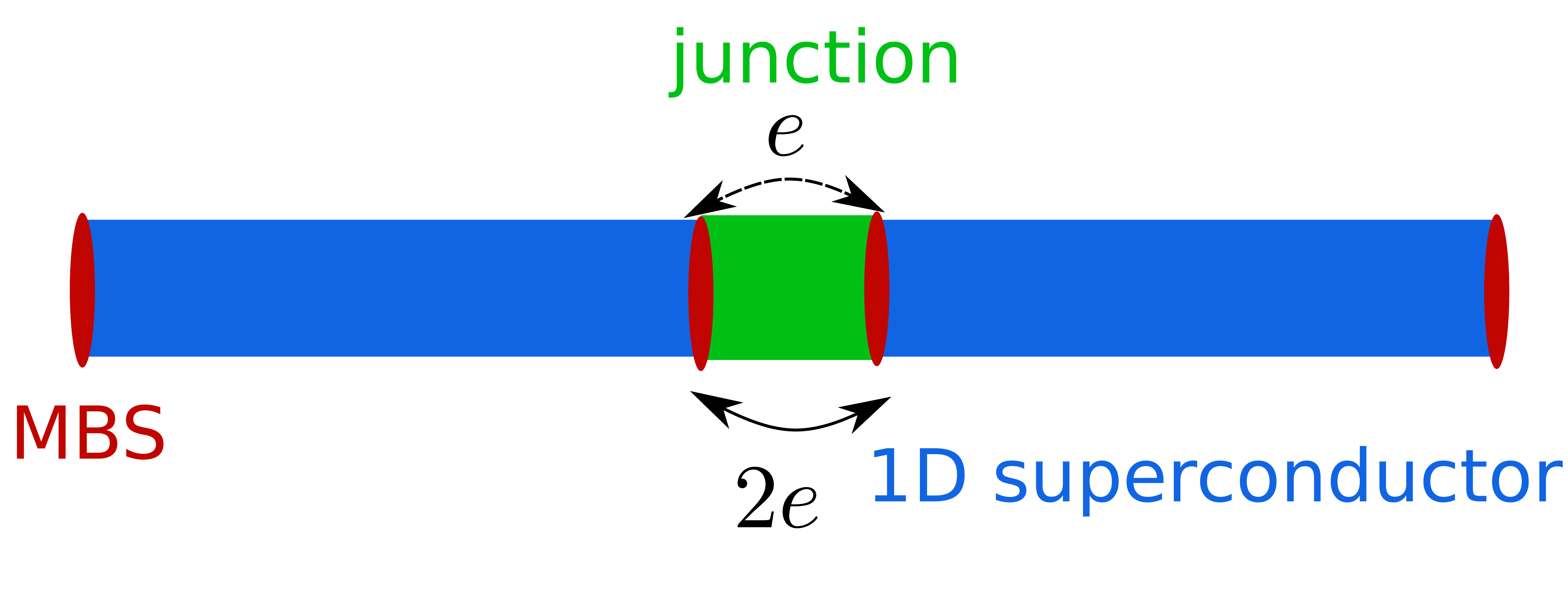}
\caption{A schematic representation of a topological JJ; $2e$ tunneling between one-dimensional superconductors corresponds to a trivial Cooper-pair tunneling, while MBS (marked with red) can induce single-electron ($e$) tunneling through the junction.}\label{fig:setup}
\end{figure} 

In this work  we focus on a short single-channel Josephson junction (JJ) that can host MBSs in the topological phase. Different physical realizations for such JJs have been proposed~\cite{Kane2008,Kane2009,Oreg2010,Linder2011,Gueron2017,Gueron2021}. Here we will assume a JJ based on a nanowire with Rashba spin-orbit interaction (SOI) in a magnetic field $B$ parallel to the wire, resulting in the Zeeman energy $V_Z=\frac{1}{2}g\mu_BB$ ($g$ is Landé $g$-factor, $\mu_B$ is Bohr magneton). An $s$-wave superconductor induces a proximity gap $\Delta$ in the nanowire, which makes the nanowire an effective one-dimensional superconductor. One should note that this $\Delta$ in general is decreasing with the field $V_Z$ (and very strong fields destroy superconductivity). While we focus for explicit calculations on nanowires, our results are more general and also apply to other JJ platforms such as topological insulators~\cite{Kane2008,Nagaosa2009,Gueron2017} and quantum spin Hall edge systems~\cite{Kane2009}. A junction is formed in an area of nanowire not covered by the supercoductor. The system is driven in the topological phase if the Zeeman field is above the critical value, in a simplified model with Fermi level in the middle of the Zeeman gap the transition occurs at $V_Z=\Delta$~\cite{Klinovaja2012}. In the topological phase a pair of MBSs, localized on the junction sides, creates an effective channel for single-quasiparticle tunneling, see Fig.~\ref{fig:setup}. As a result, this tunneling is usually expected to be characterized by the energy scale $E_M\sim\sqrt{D_N}\Delta_c$~\cite{Kwon2004,Kane2009}, corresponding to the overlap of the MBSs on the JJ, $D_N$ is the normal state transmission of the JJ, and $\Delta_c$ is the topological bulk gap. As long as we can neglect the overlap with another pair of MBSs, localized on the outer sides of the topological part of the system, the only mechanism to change the parity of the state formed by MBSs on the JJ is to absorb a quasiparticle. The parity switching events occur on time scales much shorter than any other process in the system, which allows us to work in a fixed-parity regime in between such switching events. The process is usually treated by a Fokker-Planck equation~\cite{Alicea2014,Recher2020}; we can consider the parity switch to be a random event with characteristic time scale $\tau_{qp}$ between two events. Between the switching events we can use the fixed-parity model. Topological JJs differ from a non-topological JJs by a $4\pi$-periodic term in the Hamiltonian, corresponding to single-electron tunneling~\cite{Kwon2004,Kane2009}. If the junction is biased by an electrical current $I$, the Hamiltonian of the system takes the form (we put $\hbar=1$ throughout)
\begin{align}
&H=\frac{q^2}{2C}+U(\phi),\label{eq:hamiltonian}\\
&U(\phi)=-U_{2\pi}(\phi)-U_{4\pi}(\phi)-I\phi/(2e),\label{eq:pp}
\end{align}
where $U(\phi)$ is the effective phase potential, consisting of $U_{2\pi}(\phi)$ and $U_{4\pi}(\phi)$ corresponding to a Cooper pair and a single electron (due to MBSs) tunneling, respectively, and bias current contribution; $C$ is the capacity of the JJ, $q$ is the charge accumulated on the junction, $e$ is the absolute value of the electron charge. The sign of the $4\pi$-periodic term is determined by the fermionic parity~\cite{Kane2009,Lutchyn2010}. In this work we focus on a specific case: we consider the regime of well defined phase $\phi$, which requires tunneling terms to dominate over Coulomb interactions, i.e. $E_J+E_M\gg{E}_c=\frac{e^2}{2C}$. Another assumption is that the junction is short (with the length $l\ll\xi$, where $\xi$ is the superconducting coherence length) with relatively small cross-section $S\ll\xi^2$. The Hamiltonian $H$ defined  in Eq. (\ref{eq:hamiltonian}) corresponds to an ideal system without dissipation. To include dissipation into consideration we use the RSJ model~\cite{Likharev1985,Tinkham1996}, which allows us to treat the junction as an ideal junction shunted by a resistance $R$, which is defined by quasiparticle current and depends explicitly on the experimental setup. We  assume an overdamped JJ regime $R\ll R_Q$ as experimentally relevant, where $R_Q=2\pi/(2e)^2$ is the resistance quantum.

In a trivial JJ, $U_{2\pi}(\phi)$ is determined by the sum of ABSs' energies, which are formed in the JJ, and continuum contribution (which is usually absent without magnetic field). The simple cosine phase dependencies of the ABS energies are not general, but valid for a JJ with low normal-state transparency. Nevertheless, the phase periodicity is general: $2\pi$-periodicity for the trivial  and $4\pi$-periodicity for the topological phase. As long as the occupation of the bound states is unchanged, one can treat the sum of these terms as the phase potential. Whenever the occupation of one bound states changes, the phase dependences in the phase potential change, too, which may have a significant effect on the phase evolution.

{\it Parity switching.}
Quasiparticles are created through Cooper pair breaking in the superconductor. The effect is crucial for the stability of topological qubits formed by MBSs; several approaches to estimate it were proposed~\cite{Visser2011,Chamon2011,Rainis2012,Trauzettel2012,Freedman2017,Nayak2018,Pikulin2021}, however, universal and experimentally verified estimates still do not exist. It was shown that upon creation quasiparticles rapidly relax to the continuum edge, while the further process of recombination into Cooper pairs or trapping into  subgap states are relatively slower~\cite{Martinis2009,Prober2004}. For temperatures sufficiently lower than the superconducting critical temperature the creation of quasiparticles due to thermal fluctuations~\cite{Langenberg1976} is exponentially suppressed and can be neglected. However, several experiments have reported quasiparticle density saturation at low temperatures~\cite{Devoret2004,Schreier2008,Martinis2009,Echternach2008}, which indicates other sources for quasiparticle creation. The observed densities of quasiparticles allow one to estimate the QP  rates in various systems; for semiconducting nanowires proximitized by a bulk superconductor the estimation shows high decoherence rates~\cite{Rainis2012}: $\Gamma_{qp}\gtrsim1-10\,\mathrm{MHz}$. It was shown experimentally that QP  in different superconducting qubits may soon get the dominant source of contribution for decoherence, as the techniques to reduce decoherence from other sources are being developed~\cite{Devoret2012,Ustinov2021,Martinis2021,Pekola2021}. In combination with experimental evidence~\cite{Ustinov2021,Oliver2020} it seems that the main source of quasiparticles at low temperatures is radioactivity, specifically radioactive muons  and/or gamma rays. The mechanism of quasiparticle bursts created by muons was theoretically described in~\cite{Nazarov2016}, later optimistic estimations have been made~\cite{Pikulin2021} assuming that for small enough devices ($V=10\mu\mathrm{m}\times200\mathrm{nm}\times10\mathrm{nm}=2\cdot10^{-2}\mu\mathrm{m}^3$) the effect can be neglected ($\Gamma^{-1}_{qp}\sim10\,\mathrm{days}$); however, the estimates rather contradict the experimentally observed values of QP rates~\cite{Martinis2009,Pekola2012} for similar-size superconducting devices \footnote{It is plausible that the main reason for this discrepancy is that the entire setup, including substrate and connections, contributes to the muon absorption, while the resulting  phonons can easily propagate from anywhere to the superconducting part of the setup where they create quasiparticle excitations~\cite{Ustinov2021,Martinis2021}. 
 However, this suggestion requires further investigation, especially on the specific samples which are expected to host MBSs.}.

\begin{figure}[h]
\includegraphics[width=8cm]{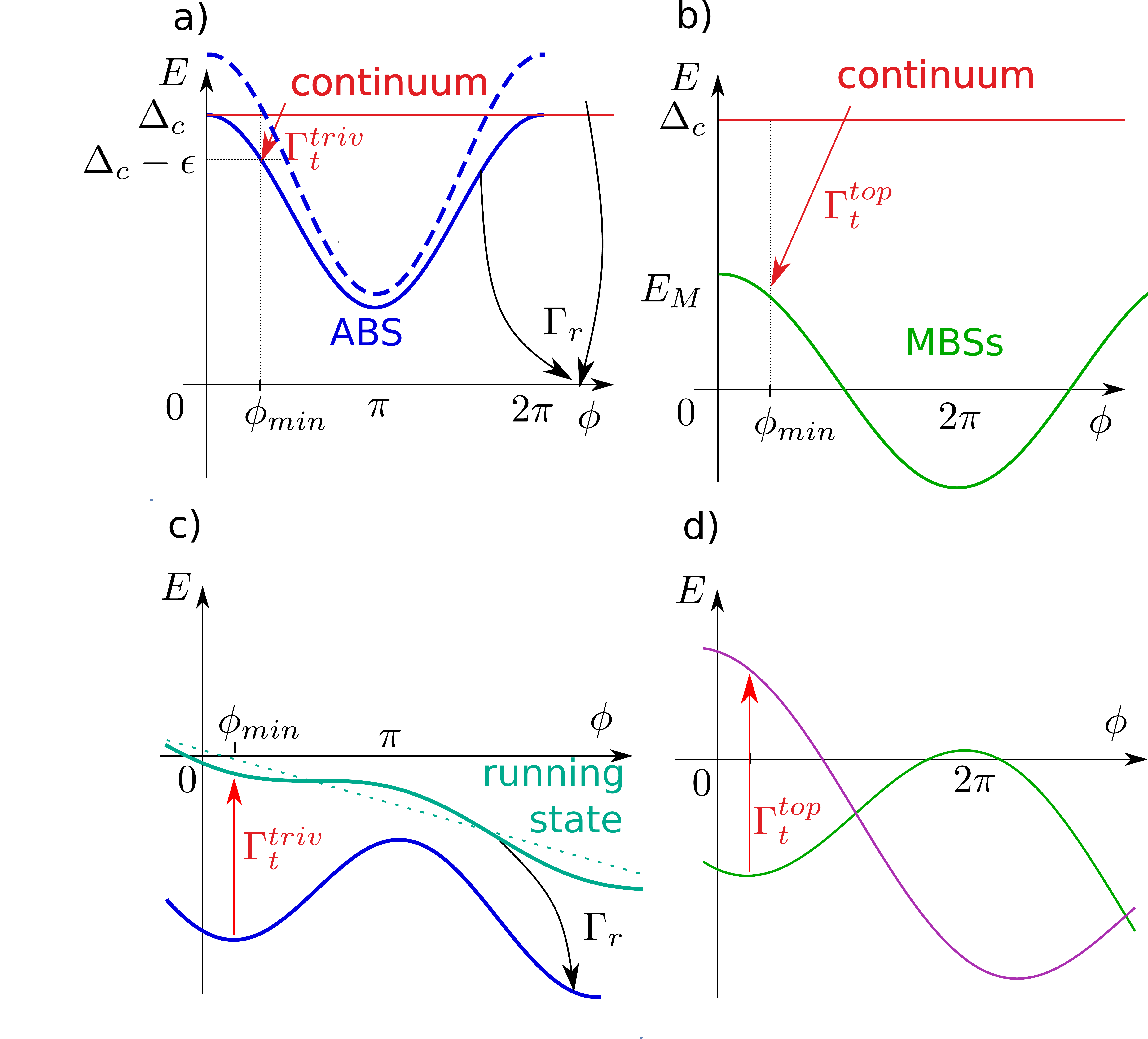}
\caption{A schematic of the single-particle spectrum of the JJ in (a) trivial  and (b) topological phase. The red solid line represents the continuum edge $\Delta_c$. Above-gap quasiparticles can get trapped in the bound states with the rate $\Gamma_t^{triv}$ in the trivial  and $\Gamma_t^{top}$ in the topological phase. When one quasiparticle is trapped in the lower ABS [solid blue curve in (a); dashed blue curve represents the higher ABS partially merged with the continuum], it can recombine with a quasiparticle from the continuum with probability $\Gamma_r\gg\Gamma_t^{triv}$. The second row corresponds to the many-particle spectrum in (c) trivial and (d) topological phase. In the trivial phase (c) trapping a quasiparticle in an ABS excites the system from the ground state [blue curve] to a state represented by the green line, which has no local minima for low fields [for $V_Z=0$ it is exactly linear, which is represented by the green dotted line in (c)], which results in the resistive (running) state.  In the topological phase (d) there is only one bound state for each parity, therefore, excitation due to QP results in a state with the same amplitude in phase but with shifted local minima [green curve for even state, violet for odd].}\label{fig:spectrum}
\end{figure} 

\begin{figure}[!tbh]
\includegraphics[width=8.5cm]{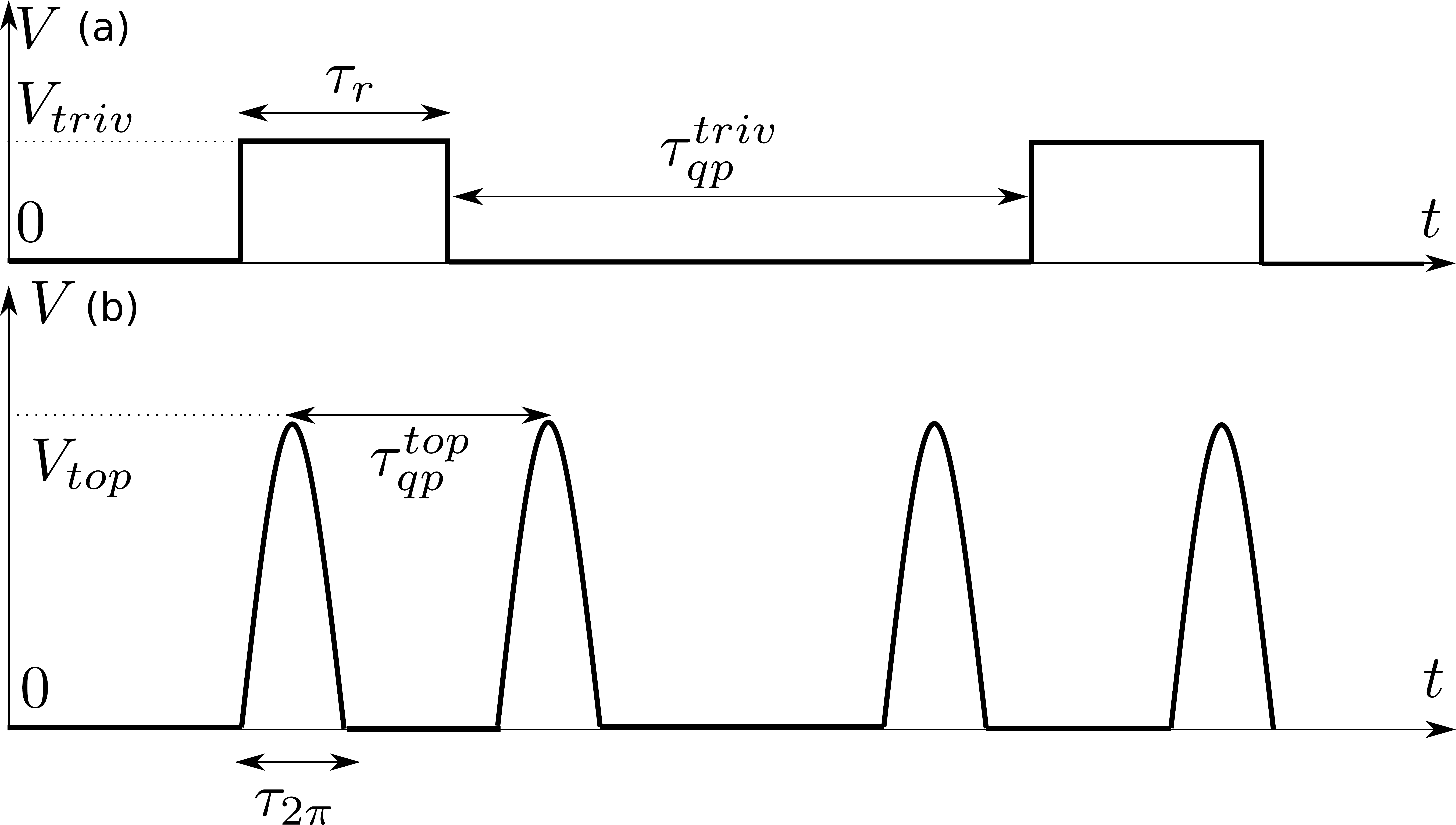}
\caption{A schematic of voltage pulses on the JJ (a) at  low magnetic fields (trivial phase) and (b) in  the topological phase. The upper panel (a) shows voltage pulses of size $V_{triv}$, given by Eq.~(\ref{eq:volttriv}), and length $\tau_r=\Gamma_r^{-1}$ and separated by $\tau_{qp}^{triv}\ll\tau_r$. These pulses correspond to the running (excited) state. In the lower panel (b) sharp voltage pulses of length $\tau_{2\pi}\ll\tau_{qp}^{top}$ and amplitude $V_{top}$ estimated by Eq.~(\ref{eq:volttop}), correspond to phase readjustment to the shifted phase potential after the parity switch. These voltage peaks are separated by $\tau_{qp}^{top}\ll\tau_{qp}^{triv}$.}\label{fig:voltage}
\end{figure}

{\it Voltage due to QP in the trivial phase.} In the trivial phase, non-equilibrium quasiparticles can recombine pairwise or relax into a subgap ABS. The latter effect is usually not easy to observe in a typical SNS-JJ with $n\gg1$ channels. If a quasiparticle gets trapped in one of the ABSs, the resulting change of the total Josephson energy is only of the order $E_J/n\ll{E}_J$, even though state-of-the-art experimental techniques allow one to observe $E_J$ fluctuations~\cite{Falk2014}. However, in a single-channel JJ, which is the system studied in this work, this argument is invalid ($n=1$). We provide a detailed analysis of the subgap spectrum in the limit of strong SOI and low transparency in the Supplemental Material (SM)~\cite{Supplemental} (for high-transparency limit see~\cite{Aguado2013,Glazman2020}). For small bias current, $I\ll I_c=2eE_J$, the minima of the phase potential are slightly shifted from the value corresponding to $I=0$: $\phi_{min}\approx 2\pi k+\frac{I}{2eE_J}$ with integer $k$, which results in a small energy gap between the continuum edge $\Delta_c$ (in the strong SOI limit $\Delta_c=\mathrm{min}\left[|\Delta-V_Z|,\Delta\right]
$) and the bound state $\epsilon\approx I^2/(4e^2E_J)\ll\Delta_c$ at $\phi_{min}$, see Fig.~\ref{fig:spectrum}. At zero magnetic field, when the ABSs are degenerate, if a quasiparticle is trapped in one ABS, the only conducting channel will get poisoned, which would result in a running (resistive) state with voltage $V=IR$. Subsequently the system relaxes back to the localized state due to recombination of the trapped quasiparticle with a quasiparticle from the continuum on a time scale $\tau_r\ll\tau_{qp}^{triv}$ (see discussion in the SM ~\cite{Supplemental} and Fig.~\ref{fig:voltage}). Upon introducing a magnetic field,  the ABSs split as well as the bulk gap $\Delta_c$ is getting smaller. As a result, the amplitude of the lower ABS is getting smaller than the higher ABS amplitude, the voltage in the running state is decreasing (quadratically in small Zeeman energy $V_Z$; moreover it is slowly varying in time) until no voltage is possible even with one quasiparticle trapped in the lowest ABS (local minima of the phase potential get restored, see Fig.~\ref{fig:spectrum} and discussion in the SM~\cite{Supplemental}). As a result, we expect that it should be possible to experimentally observe voltage pulses of length $\tau_r$ and average amplitude~\cite{Supplemental}
\begin{equation}\label{eq:volttriv}
V_{triv}\approx R\sqrt{I^{2}-e^{2}V_Z^{2}D_N^{2}/16},
\end{equation} 
separated by $\tau_{qp}^{triv}\gg\tau_{r}$ for low fields and no voltage upon approaching the topological phase transition. 

{\it Voltage due to QP in the topological phase.}
As shown for a short JJ based on a quantum spin Hall insulator~\cite{Alicea2014,Recher2020}, deep inside the topological phase, the phase potential is $4\pi$-periodic (for zero bias current) due to a fermionic state formed by two MBSs localized on the sides of the JJ, which is equivalent to $U_{2\pi}=0$ and $U_{4\pi}=(E_M/2)\cos(\phi/2)$ in Eq.~(\ref{eq:hamiltonian}). If the parity of this state is changed, the corresponding term in the Hamiltonian changes sign, which effectively flips the phase potential. Therefore, if the system is in the ground state (localized in one of the minima of the phase potential), each parity switching event should result in a time-evolution of this state, in case of sufficient dissipation it will be relaxation to the nearby new minimum. The bias current $I$ tilts the phase potential and, therefore, introduces asymmetry in the phase evolution. Then, as  mentioned in~\cite{Recher2020}, each parity switching event would result in a voltage pulse, as the phase slides by $2\pi$ to a new local minimum. The effect should be general for any JJ which can undergo a topological phase transition by varying some parameter (e.g. magnetic field). The situation is somewhat different close to the topological phase transition: in the topological phase, the phase potential is $4\pi$-periodic but has a significant $2\pi$-periodic component due to the contribution from the continuum 
states~\cite{Aguado2013,Glazman2020,Prada2020,Supplemental}, which creates additional `odd' local minima at $\phi=2\pi(2k+1)$, where $k$ is integer. The change of the bound state parity would only change the sign of the $4\pi$-periodic component, therefore, shifting `even' minima up and `odd' minima down in energy; the state would remain localized in the same local minimum of the phase potential. Nevertheless, deep inside the topological phase the $2\pi$-periodic component is suppressed ($E_M\gg E_J$) and voltage pulses would be triggered by each parity switch, if the JJ is biased by a small current. 

We illustrate this general discussion with an effective model given by Hamiltonian~(\ref{eq:hamiltonian}) with cosine phase dependences. The critical current $I_c=2e(E_J+E_M/4)$ is determined as the maximum current at which the phase potential~(\ref{eq:pp})
has local minima. If we consider the low-temperature limit with no quasiparticles, the JJ is in the perfectly superconducting state, the wave function is trapped in a local minimum of the tilted washboard potential, as long as thermal fluctuations can be neglected, i.e. $T\ll E_M$. What we are interested in is the trapping of non-equilibrium above-gap quasiparticles into the bound states on the JJ.  Deep inside the topological phase ($E_M\gg E_J$) the bound state energy alone is given by $E=E_M\cos(\phi/2)$. In the regime of strong SOI and in the low-transparency limit (high-transparency limit was studied in~\cite{Glazman2020}) we show that its amplitude scales linearly with $D_N$~\cite{Supplemental}:
\begin{equation}
E_M\approx2\frac{D_N\Delta}{V_Z}(V_Z-\Delta).
\end{equation}
This state is separated from the continuum at any phase $\phi$, which makes it prone to trap quasiparticles.
Each parity switching event effectively shifts the local minima of the phase potential by $2\pi$, causing the state to relax to this new minimum and, therefore, resulting in a voltage pulse 
\begin{equation}\label{eq:volttop}
V_{top}\approx\frac{1}{2e}\frac{2\pi}{\tau_{2\pi}} ,
\end{equation}
where $\tau_{2\pi}\approx\left(\pi E_MR/R_Q\right)^{-1}$~\cite{Beenakker2013,Alicea2014,Wiedenmann2016} is the characteristic time for relaxation in a new $2\pi$-shifted minimum calculated in the classical limit for an overdamped junction, see Fig.~\ref{fig:voltage}. The asymmetry of the phase evolution after the parity switch is determined by the bias current $I$. Then, the average voltage (averaged over many poisoning events) is given by a simple expression~\cite{Recher2020}, which does not depend on the bias current as long as it is small: 
\begin{equation}\label{eq:voltage}
\langle{V}\rangle=\frac{1}{2e}\frac{2\pi}{\tau_{qp}^{top}},
\end{equation}
where $\tau_{qp}^{top}$ is significantly shorter than $\tau_{qp}^{triv}$ in the trivial phase~\cite{Supplemental}:
\begin{equation}
\frac{\tau_{qp}^{top}}{\tau_{qp}^{triv}}\approx\frac{I^4}{(2e)^4E_J^2(\Delta_c-E_M)^2}\ll1,
\end{equation}
as the bound state is well separated from the continuum $\Delta_c-E_M\gg I/(2e)$. Eq.~(\ref{eq:voltage}) is valid as long as $\tau_{qp}^{top}<\tau_{2\pi}$, otherwise the phase does not readjust to the new minimum between the QP events; usually the phase relaxation is expected to be fast enough~\cite{Recher2020,Beenakker2013}. What is important is that we do not require a topological protection of the MBSs, more precisely, there is no need for exponential suppression of the MBSs overlap with the nanowire length, which might be hard (or even impossible) to realize due to phonons at any finite (phonon) temperature~\cite{Aseev2019}. If the resulting anticrossing~\cite{Dominguez2012,Pikulin2012} is sufficiently strong, the phase would slide only by $\pi$ to the new minimum at the anticrossing and the resulting average voltage is reduced by a factor of $2$ (see discussion in the SM~\cite{Supplemental}).

{\it Measurement.} An important question is whether the discussed effect can be measured in a realistic experimental setup. First, there could be problems with measuring the average voltage in the topological phase due to QP, since in an ideal case one can distinguish a voltage of the order of hundred nanovolts, which requires the average time between the poisoning events to be of the order of ten nanoseconds or less to see the voltage given by Eq.~(\ref{eq:voltage}). 
On the other hand, it seems that state-of-the-art experimental techniques should allow one to distinguish separate voltage peaks; using the estimate provided in~\cite{Recher2020} $\tau_{2\pi}\sim10\,\mathrm{ps}$, one can expect the measurement itself to last longer, however,  measurements at frequency $\nu_m\sim1\,\mathrm{GHz}$ are feasible, which would result in a value of the order of $V_m\approx2\pi\nu_m/(2e)\sim1\,\mu\mathrm{V}$ measured for the phase shift of $2\pi$ expected after the parity switch. The voltage pulses in the topological phase are independent of the bias current in contrast to the ones in the trivial phase. Moreover, the voltage appears in the trivial phase only for low magnetic fields (parametrically low in $I/I_c$). Finally, we expect that the characteristic time between QP events in the topological phase is much shorter than in the trivial phase $\tau_{qp}^{top}\ll\tau_{qp}^{triv}$. As a result, voltage pulses in the topological phase are more frequent, with much larger amplitude but shorter duration.

{\it Conclusions.} 
In summary, we have proposed a set of experiments to observe the effects of QP in trivial and topological JJs. Successful observation of current-independent voltage pulses in the topological phase would be a reliable way to distinguish MBSs from ABSs. Moreover, measurements in both trivial and topological phase would allow one to estimate the rates of QP in each phase. The measurement in the trivial phase could be performed right away on the existing single-channel junctions. Furthermore, performing experiments in different setups could be a way to test different sources of QP and develop efficient methods of protection against QP. 

{\it Acknowledgements.} We want to thank V. Kurilovich, D. Lvov, and D. Antonenko for fruitful discussions. This work was supported by the Swiss National Science Foundation and NCCR QSIT.
This project received funding from the European Union's Horizon 2020 research and innovation program (ERC Starting Grant, grant agreement No 757725). 
\bibliographystyle{apsrev4-1}
\bibliography{QP}

\begin{thebibliography}{46}%
\makeatletter
\providecommand \@ifxundefined [1]{%
 \@ifx{#1\undefined}
}%
\providecommand \@ifnum [1]{%
 \ifnum #1\expandafter \@firstoftwo
 \else \expandafter \@secondoftwo
 \fi
}%
\providecommand \@ifx [1]{%
 \ifx #1\expandafter \@firstoftwo
 \else \expandafter \@secondoftwo
 \fi
}%
\providecommand \natexlab [1]{#1}%
\providecommand \enquote  [1]{``#1''}%
\providecommand \bibnamefont  [1]{#1}%
\providecommand \bibfnamefont [1]{#1}%
\providecommand \citenamefont [1]{#1}%
\providecommand \href@noop [0]{\@secondoftwo}%
\providecommand \href [0]{\begingroup \@sanitize@url \@href}%
\providecommand \@href[1]{\@@startlink{#1}\@@href}%
\providecommand \@@href[1]{\endgroup#1\@@endlink}%
\providecommand \@sanitize@url [0]{\catcode `\\12\catcode `\$12\catcode
  `\&12\catcode `\#12\catcode `\^12\catcode `\_12\catcode `\%12\relax}%
\providecommand \@@startlink[1]{}%
\providecommand \@@endlink[0]{}%
\providecommand \url  [0]{\begingroup\@sanitize@url \@url }%
\providecommand \@url [1]{\endgroup\@href {#1}{\urlprefix }}%
\providecommand \urlprefix  [0]{URL }%
\providecommand \Eprint [0]{\href }%
\providecommand \doibase [0]{http://dx.doi.org/}%
\providecommand \selectlanguage [0]{\@gobble}%
\providecommand \bibinfo  [0]{\@secondoftwo}%
\providecommand \bibfield  [0]{\@secondoftwo}%
\providecommand \translation [1]{[#1]}%
\providecommand \BibitemOpen [0]{}%
\providecommand \bibitemStop [0]{}%
\providecommand \bibitemNoStop [0]{.\EOS\space}%
\providecommand \EOS [0]{\spacefactor3000\relax}%
\providecommand \BibitemShut  [1]{\csname bibitem#1\endcsname}%
\let\auto@bib@innerbib\@empty
%</preamble>
\bibitem [{\citenamefont {Kitaev}(2001)}]{Kitaev2001}%
  \BibitemOpen
  \bibfield  {author} {\bibinfo {author} {\bibfnamefont {A.~Y.}\ \bibnamefont
  {Kitaev}},\ }\href@noop {} {\bibfield  {journal} {\bibinfo  {journal}
  {Physics-Uspekhi}\ }\textbf {\bibinfo {volume} {44}},\ \bibinfo {pages} {131}
  (\bibinfo {year} {2001})}\BibitemShut {NoStop}%
\bibitem [{\citenamefont {Kwon}\ \emph {et~al.}(2003)\citenamefont {Kwon},
  \citenamefont {Sengupta},\ and\ \citenamefont {Yakovenko}}]{Kwon2003}%
  \BibitemOpen
  \bibfield  {author} {\bibinfo {author} {\bibfnamefont {H.-J.}\ \bibnamefont
  {Kwon}}, \bibinfo {author} {\bibfnamefont {K.}~\bibnamefont {Sengupta}}, \
  and\ \bibinfo {author} {\bibfnamefont {V.~M.}\ \bibnamefont {Yakovenko}},\
  }\href@noop {} {\bibfield  {journal} {\bibinfo  {journal} {Eur. Phys. J. B}\
  }\textbf {\bibinfo {volume} {37}},\ \bibinfo {pages} {349} (\bibinfo {year}
  {2003})}\BibitemShut {NoStop}%
\bibitem [{\citenamefont {Fu}\ and\ \citenamefont {Kane}(2009)}]{Kane2009}%
  \BibitemOpen
  \bibfield  {author} {\bibinfo {author} {\bibfnamefont {L.}~\bibnamefont
  {Fu}}\ and\ \bibinfo {author} {\bibfnamefont {C.~L.}\ \bibnamefont {Kane}},\
  }\href@noop {} {\bibfield  {journal} {\bibinfo  {journal} {Phys. Rev. B}\
  }\textbf {\bibinfo {volume} {79}},\ \bibinfo {pages} {161408} (\bibinfo
  {year} {2009})}\BibitemShut {NoStop}%
\bibitem [{\citenamefont {Wiedenmann}\ \emph {et~al.}(2016)\citenamefont
  {Wiedenmann}, \citenamefont {Bocquillon}, \citenamefont {Deacon},
  \citenamefont {Hartinger}, \citenamefont {Herrmann}, \citenamefont
  {Klapwijk}, \citenamefont {Maier}, \citenamefont {Ames}, \citenamefont
  {Gould}, \citenamefont {Oiwa}, \citenamefont {Ishibashi}, \citenamefont
  {Tarucha}, \citenamefont {Buhmann},\ and\ \citenamefont
  {Molenkamp}}]{Wiedenmann2016}%
  \BibitemOpen
  \bibfield  {author} {\bibinfo {author} {\bibfnamefont {J.}~\bibnamefont
  {Wiedenmann}}, \bibinfo {author} {\bibfnamefont {E.}~\bibnamefont
  {Bocquillon}}, \bibinfo {author} {\bibfnamefont {R.}~\bibnamefont {Deacon}},
  \bibinfo {author} {\bibfnamefont {S.}~\bibnamefont {Hartinger}}, \bibinfo
  {author} {\bibfnamefont {O.}~\bibnamefont {Herrmann}}, \bibinfo {author}
  {\bibfnamefont {T.~M.}\ \bibnamefont {Klapwijk}}, \bibinfo {author}
  {\bibfnamefont {L.}~\bibnamefont {Maier}}, \bibinfo {author} {\bibfnamefont
  {C.}~\bibnamefont {Ames}, \bibfnamefont {C.~Brüne}}, \bibinfo {author}
  {\bibfnamefont {C.}~\bibnamefont {Gould}}, \bibinfo {author} {\bibfnamefont
  {A.}~\bibnamefont {Oiwa}}, \bibinfo {author} {\bibfnamefont {K.}~\bibnamefont
  {Ishibashi}}, \bibinfo {author} {\bibfnamefont {S.}~\bibnamefont {Tarucha}},
  \bibinfo {author} {\bibfnamefont {H.}~\bibnamefont {Buhmann}}, \ and\
  \bibinfo {author} {\bibfnamefont {L.~W.}\ \bibnamefont {Molenkamp}},\
  }\href@noop {} {\bibfield  {journal} {\bibinfo  {journal} {Nat. Commun.}\
  }\textbf {\bibinfo {volume} {7}},\ \bibinfo {pages} {10303} (\bibinfo {year}
  {2016})}\BibitemShut {NoStop}%
\bibitem [{\citenamefont {de~Visser}\ \emph {et~al.}(2011)\citenamefont
  {de~Visser}, \citenamefont {Baselmans}, \citenamefont {Diener}, \citenamefont
  {Yates}, \citenamefont {Endo},\ and\ \citenamefont {Klapwijk}}]{Visser2011}%
  \BibitemOpen
  \bibfield  {author} {\bibinfo {author} {\bibfnamefont {P.~J.}\ \bibnamefont
  {de~Visser}}, \bibinfo {author} {\bibfnamefont {J.~J.~A.}\ \bibnamefont
  {Baselmans}}, \bibinfo {author} {\bibfnamefont {P.}~\bibnamefont {Diener}},
  \bibinfo {author} {\bibfnamefont {S.~J.~C.}\ \bibnamefont {Yates}}, \bibinfo
  {author} {\bibfnamefont {A.}~\bibnamefont {Endo}}, \ and\ \bibinfo {author}
  {\bibfnamefont {T.~M.}\ \bibnamefont {Klapwijk}},\ }\href@noop {} {\bibfield
  {journal} {\bibinfo  {journal} {Phys. Rev. Lett.}\ }\textbf {\bibinfo
  {volume} {106}},\ \bibinfo {pages} {167004} (\bibinfo {year}
  {2011})}\BibitemShut {NoStop}%
\bibitem [{\citenamefont {Goldstein}\ and\ \citenamefont
  {Chamon}(2011)}]{Chamon2011}%
  \BibitemOpen
  \bibfield  {author} {\bibinfo {author} {\bibfnamefont {G.}~\bibnamefont
  {Goldstein}}\ and\ \bibinfo {author} {\bibfnamefont {C.}~\bibnamefont
  {Chamon}},\ }\href@noop {} {\bibfield  {journal} {\bibinfo  {journal} {Phys.
  Rev. B}\ }\textbf {\bibinfo {volume} {84}},\ \bibinfo {pages} {205109}
  (\bibinfo {year} {2011})}\BibitemShut {NoStop}%
\bibitem [{\citenamefont {Rainis}\ and\ \citenamefont
  {Loss}(2012)}]{Rainis2012}%
  \BibitemOpen
  \bibfield  {author} {\bibinfo {author} {\bibfnamefont {D.}~\bibnamefont
  {Rainis}}\ and\ \bibinfo {author} {\bibfnamefont {D.}~\bibnamefont {Loss}},\
  }\href@noop {} {\bibfield  {journal} {\bibinfo  {journal} {Phys. Rev. B}\
  }\textbf {\bibinfo {volume} {85}},\ \bibinfo {pages} {174533} (\bibinfo
  {year} {2012})}\BibitemShut {NoStop}%
\bibitem [{\citenamefont {Budich}\ \emph {et~al.}(2012)\citenamefont {Budich},
  \citenamefont {Walter},\ and\ \citenamefont {Trauzettel}}]{Trauzettel2012}%
  \BibitemOpen
  \bibfield  {author} {\bibinfo {author} {\bibfnamefont {J.~C.}\ \bibnamefont
  {Budich}}, \bibinfo {author} {\bibfnamefont {S.}~\bibnamefont {Walter}}, \
  and\ \bibinfo {author} {\bibfnamefont {B.}~\bibnamefont {Trauzettel}},\
  }\href@noop {} {\bibfield  {journal} {\bibinfo  {journal} {Phys. Rev. B.}\
  }\textbf {\bibinfo {volume} {85}},\ \bibinfo {pages} {121405(R)} (\bibinfo
  {year} {2012})}\BibitemShut {NoStop}%
\bibitem [{\citenamefont {Karzig}\ \emph {et~al.}(2017)\citenamefont {Karzig},
  \citenamefont {Knapp}, \citenamefont {Lutchyn}, \citenamefont {Bonderson},
  \citenamefont {Hastings}, \citenamefont {Nayak}, \citenamefont {Alicea},
  \citenamefont {Flensberg}, \citenamefont {Plugge}, \citenamefont {Oreg},
  \citenamefont {Marcus},\ and\ \citenamefont {Freedman}}]{Freedman2017}%
  \BibitemOpen
  \bibfield  {author} {\bibinfo {author} {\bibfnamefont {T.}~\bibnamefont
  {Karzig}}, \bibinfo {author} {\bibfnamefont {C.}~\bibnamefont {Knapp}},
  \bibinfo {author} {\bibfnamefont {R.~M.}\ \bibnamefont {Lutchyn}}, \bibinfo
  {author} {\bibfnamefont {P.}~\bibnamefont {Bonderson}}, \bibinfo {author}
  {\bibfnamefont {M.~B.}\ \bibnamefont {Hastings}}, \bibinfo {author}
  {\bibfnamefont {C.}~\bibnamefont {Nayak}}, \bibinfo {author} {\bibfnamefont
  {J.}~\bibnamefont {Alicea}}, \bibinfo {author} {\bibfnamefont
  {K.}~\bibnamefont {Flensberg}}, \bibinfo {author} {\bibfnamefont
  {S.}~\bibnamefont {Plugge}}, \bibinfo {author} {\bibfnamefont
  {Y.}~\bibnamefont {Oreg}}, \bibinfo {author} {\bibfnamefont {C.~M.}\
  \bibnamefont {Marcus}}, \ and\ \bibinfo {author} {\bibfnamefont {M.~H.}\
  \bibnamefont {Freedman}},\ }\href@noop {} {\bibfield  {journal} {\bibinfo
  {journal} {Phys. Rev. B}\ }\textbf {\bibinfo {volume} {95}},\ \bibinfo
  {pages} {235305} (\bibinfo {year} {2017})}\BibitemShut {NoStop}%
\bibitem [{\citenamefont {Knapp}\ \emph {et~al.}(2018)\citenamefont {Knapp},
  \citenamefont {Karzig}, \citenamefont {Lutchyn},\ and\ \citenamefont
  {Nayak}}]{Nayak2018}%
  \BibitemOpen
  \bibfield  {author} {\bibinfo {author} {\bibfnamefont {C.}~\bibnamefont
  {Knapp}}, \bibinfo {author} {\bibfnamefont {T.}~\bibnamefont {Karzig}},
  \bibinfo {author} {\bibfnamefont {R.~M.}\ \bibnamefont {Lutchyn}}, \ and\
  \bibinfo {author} {\bibfnamefont {C.}~\bibnamefont {Nayak}},\ }\href@noop {}
  {\bibfield  {journal} {\bibinfo  {journal} {Phys. Rev. B}\ }\textbf {\bibinfo
  {volume} {97}},\ \bibinfo {pages} {125404} (\bibinfo {year}
  {2018})}\BibitemShut {NoStop}%
\bibitem [{\citenamefont {Karzig}\ \emph {et~al.}(2021)\citenamefont {Karzig},
  \citenamefont {Cole},\ and\ \citenamefont {Pikulin}}]{Pikulin2021}%
  \BibitemOpen
  \bibfield  {author} {\bibinfo {author} {\bibfnamefont {T.}~\bibnamefont
  {Karzig}}, \bibinfo {author} {\bibfnamefont {W.~S.}\ \bibnamefont {Cole}}, \
  and\ \bibinfo {author} {\bibfnamefont {D.~I.}\ \bibnamefont {Pikulin}},\
  }\href@noop {} {\bibfield  {journal} {\bibinfo  {journal} {Phys. Rev. Lett.}\
  }\textbf {\bibinfo {volume} {126}},\ \bibinfo {pages} {057702} (\bibinfo
  {year} {2021})}\BibitemShut {NoStop}%
\bibitem [{\citenamefont {Fu}\ and\ \citenamefont {Kane}(2008)}]{Kane2008}%
  \BibitemOpen
  \bibfield  {author} {\bibinfo {author} {\bibfnamefont {L.}~\bibnamefont
  {Fu}}\ and\ \bibinfo {author} {\bibfnamefont {C.~L.}\ \bibnamefont {Kane}},\
  }\href@noop {} {\bibfield  {journal} {\bibinfo  {journal} {Phys. Rev. Lett.}\
  }\textbf {\bibinfo {volume} {100}},\ \bibinfo {pages} {096407} (\bibinfo
  {year} {2008})}\BibitemShut {NoStop}%
\bibitem [{\citenamefont {Oreg}\ \emph {et~al.}(2010)\citenamefont {Oreg},
  \citenamefont {Refael},\ and\ \citenamefont {von Oppen}}]{Oreg2010}%
  \BibitemOpen
  \bibfield  {author} {\bibinfo {author} {\bibfnamefont {Y.}~\bibnamefont
  {Oreg}}, \bibinfo {author} {\bibfnamefont {G.}~\bibnamefont {Refael}}, \ and\
  \bibinfo {author} {\bibfnamefont {F.}~\bibnamefont {von Oppen}},\ }\href@noop
  {} {\bibfield  {journal} {\bibinfo  {journal} {Phys. Rev. Lett.}\ }\textbf
  {\bibinfo {volume} {105}},\ \bibinfo {pages} {177002} (\bibinfo {year}
  {2010})}\BibitemShut {NoStop}%
\bibitem [{\citenamefont {Black-Schaffer}\ and\ \citenamefont
  {Linder}(2011)}]{Linder2011}%
  \BibitemOpen
  \bibfield  {author} {\bibinfo {author} {\bibfnamefont {A.~M.}\ \bibnamefont
  {Black-Schaffer}}\ and\ \bibinfo {author} {\bibfnamefont {J.}~\bibnamefont
  {Linder}},\ }\href@noop {} {\bibfield  {journal} {\bibinfo  {journal} {Phys.
  Rev. B}\ }\textbf {\bibinfo {volume} {83}},\ \bibinfo {pages} {220511(R)}
  (\bibinfo {year} {2011})}\BibitemShut {NoStop}%
\bibitem [{\citenamefont {Murani}\ \emph {et~al.}(2017)\citenamefont {Murani},
  \citenamefont {Kasumov}, \citenamefont {Sengupta}, \citenamefont {Kasumov},
  \citenamefont {Volkov}, \citenamefont {Khodos}, \citenamefont {Brisset},
  \citenamefont {Delagrange}, \citenamefont {Chepelianskii}, \citenamefont
  {Deblock}, \citenamefont {Bouchiat},\ and\ \citenamefont
  {Gu\'eron}}]{Gueron2017}%
  \BibitemOpen
  \bibfield  {author} {\bibinfo {author} {\bibfnamefont {A.}~\bibnamefont
  {Murani}}, \bibinfo {author} {\bibfnamefont {A.}~\bibnamefont {Kasumov}},
  \bibinfo {author} {\bibfnamefont {S.}~\bibnamefont {Sengupta}}, \bibinfo
  {author} {\bibfnamefont {Y.~A.}\ \bibnamefont {Kasumov}}, \bibinfo {author}
  {\bibfnamefont {V.~T.}\ \bibnamefont {Volkov}}, \bibinfo {author}
  {\bibfnamefont {I.~I.}\ \bibnamefont {Khodos}}, \bibinfo {author}
  {\bibfnamefont {F.}~\bibnamefont {Brisset}}, \bibinfo {author} {\bibfnamefont
  {R.}~\bibnamefont {Delagrange}}, \bibinfo {author} {\bibfnamefont
  {A.}~\bibnamefont {Chepelianskii}}, \bibinfo {author} {\bibfnamefont
  {R.}~\bibnamefont {Deblock}}, \bibinfo {author} {\bibfnamefont
  {H.}~\bibnamefont {Bouchiat}}, \ and\ \bibinfo {author} {\bibfnamefont
  {S.}~\bibnamefont {Gu\'eron}},\ }\href@noop {} {\bibfield  {journal}
  {\bibinfo  {journal} {Natt. Comm.}\ }\textbf {\bibinfo {volume} {8}},\
  \bibinfo {pages} {15941} (\bibinfo {year} {2017})}\BibitemShut {NoStop}%
\bibitem [{\citenamefont {Bernard}\ \emph {et~al.}(2021)\citenamefont
  {Bernard}, \citenamefont {Kasumov}, \citenamefont {Deblock}, \citenamefont
  {Ferrier}, \citenamefont {Fortuna}, \citenamefont {Volkov}, \citenamefont
  {Kasumov}, \citenamefont {Oreg}, \citenamefont {von Oppen}, \citenamefont
  {Bouchiat},\ and\ \citenamefont {Gu\'eron}}]{Gueron2021}%
  \BibitemOpen
  \bibfield  {author} {\bibinfo {author} {\bibfnamefont {Y.}~\bibnamefont
  {Bernard}, \bibfnamefont {A.~andPeng}}, \bibinfo {author} {\bibfnamefont
  {A.}~\bibnamefont {Kasumov}}, \bibinfo {author} {\bibfnamefont
  {R.}~\bibnamefont {Deblock}}, \bibinfo {author} {\bibfnamefont
  {M.}~\bibnamefont {Ferrier}}, \bibinfo {author} {\bibfnamefont
  {F.}~\bibnamefont {Fortuna}}, \bibinfo {author} {\bibfnamefont {V.~T.}\
  \bibnamefont {Volkov}}, \bibinfo {author} {\bibfnamefont {Y.~A.}\
  \bibnamefont {Kasumov}}, \bibinfo {author} {\bibfnamefont {Y.}~\bibnamefont
  {Oreg}}, \bibinfo {author} {\bibfnamefont {F.}~\bibnamefont {von Oppen}},
  \bibinfo {author} {\bibfnamefont {H.}~\bibnamefont {Bouchiat}}, \ and\
  \bibinfo {author} {\bibfnamefont {S.}~\bibnamefont {Gu\'eron}},\ }\href@noop
  {} {\bibfield  {journal} {\bibinfo  {journal} {arXiv:2110.13539}\ } (\bibinfo
  {year} {2021})}\BibitemShut {NoStop}%
\bibitem [{\citenamefont {Tanaka}\ \emph {et~al.}(2009)\citenamefont {Tanaka},
  \citenamefont {Yokoyama},\ and\ \citenamefont {Nagaosa}}]{Nagaosa2009}%
  \BibitemOpen
  \bibfield  {author} {\bibinfo {author} {\bibfnamefont {Y.}~\bibnamefont
  {Tanaka}}, \bibinfo {author} {\bibfnamefont {T.}~\bibnamefont {Yokoyama}}, \
  and\ \bibinfo {author} {\bibfnamefont {N.}~\bibnamefont {Nagaosa}},\
  }\href@noop {} {\bibfield  {journal} {\bibinfo  {journal} {Phys. Rev. Lett.}\
  }\textbf {\bibinfo {volume} {103}},\ \bibinfo {pages} {107002} (\bibinfo
  {year} {2009})}\BibitemShut {NoStop}%
\bibitem [{\citenamefont {Klinovaja}\ and\ \citenamefont
  {Loss}(2012)}]{Klinovaja2012}%
  \BibitemOpen
  \bibfield  {author} {\bibinfo {author} {\bibfnamefont {J.}~\bibnamefont
  {Klinovaja}}\ and\ \bibinfo {author} {\bibfnamefont {D.}~\bibnamefont
  {Loss}},\ }\href@noop {} {\bibfield  {journal} {\bibinfo  {journal} {Phys.
  Rev. B}\ }\textbf {\bibinfo {volume} {86}},\ \bibinfo {pages} {085408}
  (\bibinfo {year} {2012})}\BibitemShut {NoStop}%
\bibitem [{\citenamefont {Kwon}\ \emph {et~al.}(2004)\citenamefont {Kwon},
  \citenamefont {Sengupta},\ and\ \citenamefont {Yakovenko}}]{Kwon2004}%
  \BibitemOpen
  \bibfield  {author} {\bibinfo {author} {\bibfnamefont {H.-J.}\ \bibnamefont
  {Kwon}}, \bibinfo {author} {\bibfnamefont {K.}~\bibnamefont {Sengupta}}, \
  and\ \bibinfo {author} {\bibfnamefont {V.~M.}\ \bibnamefont {Yakovenko}},\
  }\href@noop {} {\bibfield  {journal} {\bibinfo  {journal} {Low. Temp. Phys.}\
  }\textbf {\bibinfo {volume} {30}},\ \bibinfo {pages} {613} (\bibinfo {year}
  {2004})}\BibitemShut {NoStop}%
\bibitem [{\citenamefont {Lee}\ \emph {et~al.}(2014)\citenamefont {Lee},
  \citenamefont {Michaeli}, \citenamefont {Alicea},\ and\ \citenamefont
  {Yacoby}}]{Alicea2014}%
  \BibitemOpen
  \bibfield  {author} {\bibinfo {author} {\bibfnamefont {S.-P.}\ \bibnamefont
  {Lee}}, \bibinfo {author} {\bibfnamefont {K.}~\bibnamefont {Michaeli}},
  \bibinfo {author} {\bibfnamefont {J.}~\bibnamefont {Alicea}}, \ and\ \bibinfo
  {author} {\bibfnamefont {A.}~\bibnamefont {Yacoby}},\ }\href@noop {}
  {\bibfield  {journal} {\bibinfo  {journal} {Phys. Rev. Lett.}\ }\textbf
  {\bibinfo {volume} {113}},\ \bibinfo {pages} {197001} (\bibinfo {year}
  {2014})}\BibitemShut {NoStop}%
\bibitem [{\citenamefont {Frombach}\ and\ \citenamefont
  {Recher}(2020)}]{Recher2020}%
  \BibitemOpen
  \bibfield  {author} {\bibinfo {author} {\bibfnamefont {D.}~\bibnamefont
  {Frombach}}\ and\ \bibinfo {author} {\bibfnamefont {P.}~\bibnamefont
  {Recher}},\ }\href@noop {} {\bibfield  {journal} {\bibinfo  {journal} {Phys.
  Rev. B}\ }\textbf {\bibinfo {volume} {101}},\ \bibinfo {pages} {115304}
  (\bibinfo {year} {2020})}\BibitemShut {NoStop}%
\bibitem [{\citenamefont {Lutchyn}\ \emph {et~al.}(2010)\citenamefont
  {Lutchyn}, \citenamefont {Sau},\ and\ \citenamefont
  {Das~Sarma}}]{Lutchyn2010}%
  \BibitemOpen
  \bibfield  {author} {\bibinfo {author} {\bibfnamefont {R.~M.}\ \bibnamefont
  {Lutchyn}}, \bibinfo {author} {\bibfnamefont {J.~D.}\ \bibnamefont {Sau}}, \
  and\ \bibinfo {author} {\bibfnamefont {S.}~\bibnamefont {Das~Sarma}},\
  }\href@noop {} {\bibfield  {journal} {\bibinfo  {journal} {Phys. Rev. Lett.}\
  }\textbf {\bibinfo {volume} {105}},\ \bibinfo {pages} {077001} (\bibinfo
  {year} {2010})}\BibitemShut {NoStop}%
\bibitem [{\citenamefont {Likharev}\ and\ \citenamefont
  {Zorin}(1985)}]{Likharev1985}%
  \BibitemOpen
  \bibfield  {author} {\bibinfo {author} {\bibfnamefont {K.~K.}\ \bibnamefont
  {Likharev}}\ and\ \bibinfo {author} {\bibfnamefont {A.~B.}\ \bibnamefont
  {Zorin}},\ }\href@noop {} {\bibfield  {journal} {\bibinfo  {journal} {J. of
  Low Temp. Phys.}\ }\textbf {\bibinfo {volume} {59}},\ \bibinfo {pages} {347}
  (\bibinfo {year} {1985})}\BibitemShut {NoStop}%
\bibitem [{\citenamefont {Tinkham}(1996)}]{Tinkham1996}%
  \BibitemOpen
  \bibfield  {author} {\bibinfo {author} {\bibfnamefont {M.}~\bibnamefont
  {Tinkham}},\ }in\ \href@noop {} {\emph {\bibinfo {booktitle} {Introduction to
  Superconductivity}}}\ (\bibinfo  {publisher} {Second Edition, McGraw-Hill
  Book Co.},\ \bibinfo {year} {1996})\BibitemShut {NoStop}%
\bibitem [{\citenamefont {Martinis}\ \emph {et~al.}(2009)\citenamefont
  {Martinis}, \citenamefont {Ansmann},\ and\ \citenamefont
  {Aumentado}}]{Martinis2009}%
  \BibitemOpen
  \bibfield  {author} {\bibinfo {author} {\bibfnamefont {J.~M.}\ \bibnamefont
  {Martinis}}, \bibinfo {author} {\bibfnamefont {M.}~\bibnamefont {Ansmann}}, \
  and\ \bibinfo {author} {\bibfnamefont {J.}~\bibnamefont {Aumentado}},\
  }\href@noop {} {\bibfield  {journal} {\bibinfo  {journal} {Phys. Rev. Lett.}\
  }\textbf {\bibinfo {volume} {103}},\ \bibinfo {pages} {097002} (\bibinfo
  {year} {2009})}\BibitemShut {NoStop}%
\bibitem [{\citenamefont {Segall}\ \emph {et~al.}(2004)\citenamefont {Segall},
  \citenamefont {Wilson}, \citenamefont {Li}, \citenamefont {Frunzio},
  \citenamefont {Friedrich}, \citenamefont {Gaidis},\ and\ \citenamefont
  {Prober}}]{Prober2004}%
  \BibitemOpen
  \bibfield  {author} {\bibinfo {author} {\bibfnamefont {K.}~\bibnamefont
  {Segall}}, \bibinfo {author} {\bibfnamefont {C.}~\bibnamefont {Wilson}},
  \bibinfo {author} {\bibfnamefont {L.}~\bibnamefont {Li}}, \bibinfo {author}
  {\bibfnamefont {L.}~\bibnamefont {Frunzio}}, \bibinfo {author} {\bibfnamefont
  {S.}~\bibnamefont {Friedrich}}, \bibinfo {author} {\bibfnamefont {M.~C.}\
  \bibnamefont {Gaidis}}, \ and\ \bibinfo {author} {\bibfnamefont {D.~E.}\
  \bibnamefont {Prober}},\ }\href@noop {} {\bibfield  {journal} {\bibinfo
  {journal} {Phys. Rev. B}\ }\textbf {\bibinfo {volume} {70}},\ \bibinfo
  {pages} {214520} (\bibinfo {year} {2004})}\BibitemShut {NoStop}%
\bibitem [{\citenamefont {Kaplan}\ \emph {et~al.}(1976)\citenamefont {Kaplan},
  \citenamefont {Chi},\ and\ \citenamefont {Langenberg}}]{Langenberg1976}%
  \BibitemOpen
  \bibfield  {author} {\bibinfo {author} {\bibfnamefont {S.~B.}\ \bibnamefont
  {Kaplan}}, \bibinfo {author} {\bibfnamefont {C.~C.}\ \bibnamefont {Chi}}, \
  and\ \bibinfo {author} {\bibfnamefont {D.~N.}\ \bibnamefont {Langenberg}},\
  }\href@noop {} {\bibfield  {journal} {\bibinfo  {journal} {Phys. Rev. B}\
  }\textbf {\bibinfo {volume} {14}},\ \bibinfo {pages} {11} (\bibinfo {year}
  {1976})}\BibitemShut {NoStop}%
\bibitem [{\citenamefont {Aumentado}\ \emph {et~al.}(2004)\citenamefont
  {Aumentado}, \citenamefont {Keller}, \citenamefont {Martinis},\ and\
  \citenamefont {Devoret}}]{Devoret2004}%
  \BibitemOpen
  \bibfield  {author} {\bibinfo {author} {\bibfnamefont {J.}~\bibnamefont
  {Aumentado}}, \bibinfo {author} {\bibfnamefont {M.~W.}\ \bibnamefont
  {Keller}}, \bibinfo {author} {\bibfnamefont {J.~M.}\ \bibnamefont
  {Martinis}}, \ and\ \bibinfo {author} {\bibfnamefont {M.~H.}\ \bibnamefont
  {Devoret}},\ }\href@noop {} {\bibfield  {journal} {\bibinfo  {journal} {Phys.
  Rev. Lett.}\ }\textbf {\bibinfo {volume} {92}},\ \bibinfo {pages} {066802}
  (\bibinfo {year} {2004})}\BibitemShut {NoStop}%
\bibitem [{\citenamefont {Schreier}\ \emph {et~al.}(2008)\citenamefont
  {Schreier}, \citenamefont {Houck}, \citenamefont {Koch}, \citenamefont
  {Schuster}, \citenamefont {Johnson}, \citenamefont {Chow}, \citenamefont
  {Gambetta}, \citenamefont {Majer}, \citenamefont {Frunzio}, \citenamefont
  {Devoret}, \citenamefont {Girvin},\ and\ \citenamefont
  {Schoelkopf}}]{Schreier2008}%
  \BibitemOpen
  \bibfield  {author} {\bibinfo {author} {\bibfnamefont {J.~A.}\ \bibnamefont
  {Schreier}}, \bibinfo {author} {\bibfnamefont {A.~A.}\ \bibnamefont {Houck}},
  \bibinfo {author} {\bibfnamefont {J.}~\bibnamefont {Koch}}, \bibinfo {author}
  {\bibfnamefont {D.~I.}\ \bibnamefont {Schuster}}, \bibinfo {author}
  {\bibfnamefont {B.~R.}\ \bibnamefont {Johnson}}, \bibinfo {author}
  {\bibfnamefont {J.~M.}\ \bibnamefont {Chow}}, \bibinfo {author}
  {\bibfnamefont {J.~M.}\ \bibnamefont {Gambetta}}, \bibinfo {author}
  {\bibfnamefont {J.}~\bibnamefont {Majer}}, \bibinfo {author} {\bibfnamefont
  {L.}~\bibnamefont {Frunzio}}, \bibinfo {author} {\bibfnamefont {M.~H.}\
  \bibnamefont {Devoret}}, \bibinfo {author} {\bibfnamefont {S.~M.}\
  \bibnamefont {Girvin}}, \ and\ \bibinfo {author} {\bibfnamefont {R.~J.}\
  \bibnamefont {Schoelkopf}},\ }\href@noop {} {\bibfield  {journal} {\bibinfo
  {journal} {Phys. Rev. B}\ }\textbf {\bibinfo {volume} {77}},\ \bibinfo
  {pages} {180502(R)} (\bibinfo {year} {2008})}\BibitemShut {NoStop}%
\bibitem [{\citenamefont {Shaw}\ \emph {et~al.}(2008)\citenamefont {Shaw},
  \citenamefont {Lutchyn}, \citenamefont {Delsing},\ and\ \citenamefont
  {Echternach}}]{Echternach2008}%
  \BibitemOpen
  \bibfield  {author} {\bibinfo {author} {\bibfnamefont {M.~D.}\ \bibnamefont
  {Shaw}}, \bibinfo {author} {\bibfnamefont {R.~M.}\ \bibnamefont {Lutchyn}},
  \bibinfo {author} {\bibfnamefont {P.}~\bibnamefont {Delsing}}, \ and\
  \bibinfo {author} {\bibfnamefont {P.~M.}\ \bibnamefont {Echternach}},\
  }\href@noop {} {\bibfield  {journal} {\bibinfo  {journal} {Phys. Rev. B}\
  }\textbf {\bibinfo {volume} {78}},\ \bibinfo {pages} {024503} (\bibinfo
  {year} {2008})}\BibitemShut {NoStop}%
\bibitem [{\citenamefont {Sun}\ \emph {et~al.}(2012)\citenamefont {Sun},
  \citenamefont {DiCarlo}, \citenamefont {Reed}, \citenamefont {Catelani},
  \citenamefont {Bishop}, \citenamefont {Schuster}, \citenamefont {Johnson},
  \citenamefont {Yang}, \citenamefont {Frunzio}, \citenamefont {Glazman},
  \citenamefont {Devoret},\ and\ \citenamefont {Schoelkopf}}]{Devoret2012}%
  \BibitemOpen
  \bibfield  {author} {\bibinfo {author} {\bibfnamefont {L.}~\bibnamefont
  {Sun}}, \bibinfo {author} {\bibfnamefont {L.}~\bibnamefont {DiCarlo}},
  \bibinfo {author} {\bibfnamefont {M.~D.}\ \bibnamefont {Reed}}, \bibinfo
  {author} {\bibfnamefont {G.}~\bibnamefont {Catelani}}, \bibinfo {author}
  {\bibfnamefont {L.~S.}\ \bibnamefont {Bishop}}, \bibinfo {author}
  {\bibfnamefont {D.~I.}\ \bibnamefont {Schuster}}, \bibinfo {author}
  {\bibfnamefont {B.~R.}\ \bibnamefont {Johnson}}, \bibinfo {author}
  {\bibfnamefont {G.~A.}\ \bibnamefont {Yang}}, \bibinfo {author}
  {\bibfnamefont {L.}~\bibnamefont {Frunzio}}, \bibinfo {author} {\bibfnamefont
  {L.}~\bibnamefont {Glazman}}, \bibinfo {author} {\bibfnamefont {M.~H.}\
  \bibnamefont {Devoret}}, \ and\ \bibinfo {author} {\bibfnamefont {R.~J.}\
  \bibnamefont {Schoelkopf}},\ }\href@noop {} {\bibfield  {journal} {\bibinfo
  {journal} {Phys. Rev. Lett.}\ }\textbf {\bibinfo {volume} {108}},\ \bibinfo
  {pages} {230509} (\bibinfo {year} {2012})}\BibitemShut {NoStop}%
\bibitem [{\citenamefont {Cardani}\ \emph {et~al.}(2021)\citenamefont
  {Cardani}, \citenamefont {Valenti}, \citenamefont {Casali}, \citenamefont
  {Catelani}, \citenamefont {Charpentier}, \citenamefont {Clemenza},
  \citenamefont {Colantoni}, \citenamefont {Cruciani}, \citenamefont
  {D’Imperio}, \citenamefont {Gironi}, \citenamefont {Grünhaupt},
  \citenamefont {Gusenkova}, \citenamefont {Henriques}, \citenamefont {Lagoin},
  \citenamefont {Martinez}, \citenamefont {Pettinari}, \citenamefont {Rusconi},
  \citenamefont {Sander}, \citenamefont {Tomei}, \citenamefont {Ustinov},
  \citenamefont {Weber}, \citenamefont {Wernsdorfer}, \citenamefont {Vignati},
  \citenamefont {S.},\ and\ \citenamefont {Pop}}]{Ustinov2021}%
  \BibitemOpen
  \bibfield  {author} {\bibinfo {author} {\bibfnamefont {L.}~\bibnamefont
  {Cardani}}, \bibinfo {author} {\bibfnamefont {F.}~\bibnamefont {Valenti}},
  \bibinfo {author} {\bibfnamefont {N.}~\bibnamefont {Casali}}, \bibinfo
  {author} {\bibfnamefont {G.}~\bibnamefont {Catelani}}, \bibinfo {author}
  {\bibfnamefont {T.}~\bibnamefont {Charpentier}}, \bibinfo {author}
  {\bibfnamefont {M.}~\bibnamefont {Clemenza}}, \bibinfo {author}
  {\bibfnamefont {I.}~\bibnamefont {Colantoni}}, \bibinfo {author}
  {\bibfnamefont {A.}~\bibnamefont {Cruciani}}, \bibinfo {author}
  {\bibfnamefont {G.}~\bibnamefont {D’Imperio}}, \bibinfo {author}
  {\bibfnamefont {L.}~\bibnamefont {Gironi}}, \bibinfo {author} {\bibfnamefont
  {L.}~\bibnamefont {Grünhaupt}}, \bibinfo {author} {\bibfnamefont
  {D.}~\bibnamefont {Gusenkova}}, \bibinfo {author} {\bibfnamefont
  {F.}~\bibnamefont {Henriques}}, \bibinfo {author} {\bibfnamefont
  {M.}~\bibnamefont {Lagoin}}, \bibinfo {author} {\bibfnamefont
  {M.}~\bibnamefont {Martinez}}, \bibinfo {author} {\bibfnamefont
  {G.}~\bibnamefont {Pettinari}}, \bibinfo {author} {\bibfnamefont
  {C.}~\bibnamefont {Rusconi}}, \bibinfo {author} {\bibfnamefont
  {O.}~\bibnamefont {Sander}}, \bibinfo {author} {\bibfnamefont
  {C.}~\bibnamefont {Tomei}}, \bibinfo {author} {\bibfnamefont {A.~V.}\
  \bibnamefont {Ustinov}}, \bibinfo {author} {\bibfnamefont {M.}~\bibnamefont
  {Weber}}, \bibinfo {author} {\bibfnamefont {W.}~\bibnamefont {Wernsdorfer}},
  \bibinfo {author} {\bibfnamefont {M.}~\bibnamefont {Vignati}}, \bibinfo
  {author} {\bibfnamefont {P.}~\bibnamefont {S.}}, \ and\ \bibinfo {author}
  {\bibfnamefont {I.~M.}\ \bibnamefont {Pop}},\ }\href@noop {} {\bibfield
  {journal} {\bibinfo  {journal} {Nat. Comm.}\ }\textbf {\bibinfo {volume}
  {12}},\ \bibinfo {pages} {2733} (\bibinfo {year} {2021})}\BibitemShut
  {NoStop}%
\bibitem [{\citenamefont {Martinis}(2021)}]{Martinis2021}%
  \BibitemOpen
  \bibfield  {author} {\bibinfo {author} {\bibfnamefont {J.~M.}\ \bibnamefont
  {Martinis}},\ }\href@noop {} {\bibfield  {journal} {\bibinfo  {journal} {npj
  Quantum Inf}\ }\textbf {\bibinfo {volume} {7}},\ \bibinfo {pages} {90}
  (\bibinfo {year} {2021})}\BibitemShut {NoStop}%
\bibitem [{\citenamefont {Catelani}\ and\ \citenamefont
  {Pekola}(2021)}]{Pekola2021}%
  \BibitemOpen
  \bibfield  {author} {\bibinfo {author} {\bibfnamefont {G.}~\bibnamefont
  {Catelani}}\ and\ \bibinfo {author} {\bibfnamefont {J.~P.}\ \bibnamefont
  {Pekola}},\ }\href@noop {} {\bibfield  {journal} {\bibinfo  {journal}
  {arXiv:2107.09695}\ } (\bibinfo {year} {2021})}\BibitemShut {NoStop}%
\bibitem [{\citenamefont {Vepsäläinen}\ \emph {et~al.}(2020)\citenamefont
  {Vepsäläinen}, \citenamefont {Karamlou}, \citenamefont {Orrell},
  \citenamefont {Dogra}, \citenamefont {Loer}, \citenamefont {Vasconcelos},
  \citenamefont {Kim}, \citenamefont {Melville}, \citenamefont {Niedzielski},
  \citenamefont {Yoder}, \citenamefont {Gustavsson}, \citenamefont {Formaggio},
  \citenamefont {VanDevender},\ and\ \citenamefont {Oliver}}]{Oliver2020}%
  \BibitemOpen
  \bibfield  {author} {\bibinfo {author} {\bibfnamefont {A.~P.}\ \bibnamefont
  {Vepsäläinen}}, \bibinfo {author} {\bibfnamefont {A.~H.}\ \bibnamefont
  {Karamlou}}, \bibinfo {author} {\bibfnamefont {J.~L.}\ \bibnamefont
  {Orrell}}, \bibinfo {author} {\bibfnamefont {A.~S.}\ \bibnamefont {Dogra}},
  \bibinfo {author} {\bibfnamefont {B.}~\bibnamefont {Loer}}, \bibinfo {author}
  {\bibfnamefont {F.}~\bibnamefont {Vasconcelos}}, \bibinfo {author}
  {\bibfnamefont {D.~K.}\ \bibnamefont {Kim}}, \bibinfo {author} {\bibfnamefont
  {A.~J.}\ \bibnamefont {Melville}}, \bibinfo {author} {\bibfnamefont {B.~M.}\
  \bibnamefont {Niedzielski}}, \bibinfo {author} {\bibfnamefont {J.~L.}\
  \bibnamefont {Yoder}}, \bibinfo {author} {\bibfnamefont {S.}~\bibnamefont
  {Gustavsson}}, \bibinfo {author} {\bibfnamefont {J.~A.}\ \bibnamefont
  {Formaggio}}, \bibinfo {author} {\bibfnamefont {B.~A.}\ \bibnamefont
  {VanDevender}}, \ and\ \bibinfo {author} {\bibfnamefont {W.~D.}\ \bibnamefont
  {Oliver}},\ }\href@noop {} {\bibfield  {journal} {\bibinfo  {journal}
  {Nature}\ }\textbf {\bibinfo {volume} {584}},\ \bibinfo {pages} {551}
  (\bibinfo {year} {2020})}\BibitemShut {NoStop}%
\bibitem [{\citenamefont {Bespalov}\ \emph {et~al.}(2016)\citenamefont
  {Bespalov}, \citenamefont {Houzet}, \citenamefont {Meyer},\ and\
  \citenamefont {Nazarov}}]{Nazarov2016}%
  \BibitemOpen
  \bibfield  {author} {\bibinfo {author} {\bibfnamefont {A.}~\bibnamefont
  {Bespalov}}, \bibinfo {author} {\bibfnamefont {M.}~\bibnamefont {Houzet}},
  \bibinfo {author} {\bibfnamefont {J.~S.}\ \bibnamefont {Meyer}}, \ and\
  \bibinfo {author} {\bibfnamefont {Y.~V.}\ \bibnamefont {Nazarov}},\
  }\href@noop {} {\bibfield  {journal} {\bibinfo  {journal} {Phys. Rev. Lett.}\
  }\textbf {\bibinfo {volume} {117}},\ \bibinfo {pages} {117002} (\bibinfo
  {year} {2016})}\BibitemShut {NoStop}%
\bibitem [{\citenamefont {Saira}\ \emph {et~al.}(2012)\citenamefont {Saira},
  \citenamefont {Kemppinen}, \citenamefont {Maisi},\ and\ \citenamefont
  {Pekola}}]{Pekola2012}%
  \BibitemOpen
  \bibfield  {author} {\bibinfo {author} {\bibfnamefont {O.-P.}\ \bibnamefont
  {Saira}}, \bibinfo {author} {\bibfnamefont {A.}~\bibnamefont {Kemppinen}},
  \bibinfo {author} {\bibfnamefont {V.~F.}\ \bibnamefont {Maisi}}, \ and\
  \bibinfo {author} {\bibfnamefont {J.~P.}\ \bibnamefont {Pekola}},\
  }\href@noop {} {\bibfield  {journal} {\bibinfo  {journal} {Phys. Rev. B}\
  }\textbf {\bibinfo {volume} {85}},\ \bibinfo {pages} {012504} (\bibinfo
  {year} {2012})}\BibitemShut {NoStop}%
\bibitem [{\citenamefont {Levenson-Falk}\ \emph {et~al.}(2014)\citenamefont
  {Levenson-Falk}, \citenamefont {Kos}, \citenamefont {Vijay}, \citenamefont
  {Glazman},\ and\ \citenamefont {Siddiqi}}]{Falk2014}%
  \BibitemOpen
  \bibfield  {author} {\bibinfo {author} {\bibfnamefont {E.~M.}\ \bibnamefont
  {Levenson-Falk}}, \bibinfo {author} {\bibfnamefont {F.}~\bibnamefont {Kos}},
  \bibinfo {author} {\bibfnamefont {R.}~\bibnamefont {Vijay}}, \bibinfo
  {author} {\bibfnamefont {L.}~\bibnamefont {Glazman}}, \ and\ \bibinfo
  {author} {\bibfnamefont {I.}~\bibnamefont {Siddiqi}},\ }\href@noop {}
  {\bibfield  {journal} {\bibinfo  {journal} {Phys. Rev. Lett.}\ }\textbf
  {\bibinfo {volume} {112}},\ \bibinfo {pages} {047002} (\bibinfo {year}
  {2014})}\BibitemShut {NoStop}%
\bibitem [{\citenamefont {Svetogorov}\ \emph {et~al.}()\citenamefont
  {Svetogorov}, \citenamefont {Loss},\ and\ \citenamefont
  {Klinovaja}}]{Supplemental}%
  \BibitemOpen
  \bibfield  {author} {\bibinfo {author} {\bibfnamefont {A.~E.}\ \bibnamefont
  {Svetogorov}}, \bibinfo {author} {\bibfnamefont {D.}~\bibnamefont {Loss}}, \
  and\ \bibinfo {author} {\bibfnamefont {J.}~\bibnamefont {Klinovaja}},\
  }\href@noop {} {\bibinfo  {journal} {Supplemental materials}\ }\BibitemShut
  {NoStop}%
\bibitem [{\citenamefont {San-Jose}\ \emph {et~al.}(2013)\citenamefont
  {San-Jose}, \citenamefont {Cayao}, \citenamefont {Prada},\ and\ \citenamefont
  {Aguado}}]{Aguado2013}%
  \BibitemOpen
\bibfield  {journal} {  }\bibfield  {author} {\bibinfo {author} {\bibfnamefont
  {P.}~\bibnamefont {San-Jose}}, \bibinfo {author} {\bibfnamefont
  {J.}~\bibnamefont {Cayao}}, \bibinfo {author} {\bibfnamefont
  {E.}~\bibnamefont {Prada}}, \ and\ \bibinfo {author} {\bibfnamefont
  {R.}~\bibnamefont {Aguado}},\ }\href@noop {} {\bibfield  {journal} {\bibinfo
  {journal} {New J. Phys.}\ }\textbf {\bibinfo {volume} {15}},\ \bibinfo
  {pages} {075019} (\bibinfo {year} {2013})}\BibitemShut {NoStop}%
\bibitem [{\citenamefont {Murthy}\ \emph {et~al.}(2020)\citenamefont {Murthy},
  \citenamefont {Kurilovich}, \citenamefont {Kurilovich}, \citenamefont {van
  Heck}, \citenamefont {Glazman},\ and\ \citenamefont {Nayak}}]{Glazman2020}%
  \BibitemOpen
  \bibfield  {author} {\bibinfo {author} {\bibfnamefont {C.}~\bibnamefont
  {Murthy}}, \bibinfo {author} {\bibfnamefont {V.~D.}\ \bibnamefont
  {Kurilovich}}, \bibinfo {author} {\bibfnamefont {P.~D.}\ \bibnamefont
  {Kurilovich}}, \bibinfo {author} {\bibfnamefont {B.}~\bibnamefont {van
  Heck}}, \bibinfo {author} {\bibfnamefont {L.~I.}\ \bibnamefont {Glazman}}, \
  and\ \bibinfo {author} {\bibfnamefont {C.}~\bibnamefont {Nayak}},\
  }\href@noop {} {\bibfield  {journal} {\bibinfo  {journal} {Phys. Rev. B.}\
  }\textbf {\bibinfo {volume} {101}},\ \bibinfo {pages} {224501} (\bibinfo
  {year} {2020})}\BibitemShut {NoStop}%
\bibitem [{\citenamefont {Prada}\ \emph {et~al.}(2020)\citenamefont {Prada},
  \citenamefont {San-Jose}, \citenamefont {de~Moor}, \citenamefont {Geresdi},
  \citenamefont {Lee}, \citenamefont {Klinovaja}, \citenamefont {Loss},
  \citenamefont {Nyg{\aa}rd}, \citenamefont {Aguado},\ and\ \citenamefont
  {Kouwenhoven}}]{Prada2020}%
  \BibitemOpen
  \bibfield  {author} {\bibinfo {author} {\bibfnamefont {E.}~\bibnamefont
  {Prada}}, \bibinfo {author} {\bibfnamefont {P.}~\bibnamefont {San-Jose}},
  \bibinfo {author} {\bibfnamefont {M.~W.~A.}\ \bibnamefont {de~Moor}},
  \bibinfo {author} {\bibfnamefont {A.}~\bibnamefont {Geresdi}}, \bibinfo
  {author} {\bibfnamefont {E.~J.~H.}\ \bibnamefont {Lee}}, \bibinfo {author}
  {\bibfnamefont {J.}~\bibnamefont {Klinovaja}}, \bibinfo {author}
  {\bibfnamefont {D.}~\bibnamefont {Loss}}, \bibinfo {author} {\bibfnamefont
  {J.}~\bibnamefont {Nyg{\aa}rd}}, \bibinfo {author} {\bibfnamefont
  {R.}~\bibnamefont {Aguado}}, \ and\ \bibinfo {author} {\bibfnamefont {L.~P.}\
  \bibnamefont {Kouwenhoven}},\ }\href@noop {} {\bibfield  {journal} {\bibinfo
  {journal} {Nature Reviews Physics}\ }\textbf {\bibinfo {volume} {2}},\
  \bibinfo {pages} {575–594} (\bibinfo {year} {2020})}\BibitemShut {NoStop}%
\bibitem [{\citenamefont {Beenakker}\ \emph {et~al.}(2013)\citenamefont
  {Beenakker}, \citenamefont {Pikulin}, \citenamefont {Hyart}, \citenamefont
  {Schomerus},\ and\ \citenamefont {Dahlhaus}}]{Beenakker2013}%
  \BibitemOpen
  \bibfield  {author} {\bibinfo {author} {\bibfnamefont {C.~W.~J.}\
  \bibnamefont {Beenakker}}, \bibinfo {author} {\bibfnamefont {D.~I.}\
  \bibnamefont {Pikulin}}, \bibinfo {author} {\bibfnamefont {T.}~\bibnamefont
  {Hyart}}, \bibinfo {author} {\bibfnamefont {H.}~\bibnamefont {Schomerus}}, \
  and\ \bibinfo {author} {\bibfnamefont {J.~P.}\ \bibnamefont {Dahlhaus}},\
  }\href@noop {} {\bibfield  {journal} {\bibinfo  {journal} {Phys. Rev. Lett.}\
  }\textbf {\bibinfo {volume} {110}},\ \bibinfo {pages} {017003} (\bibinfo
  {year} {2013})}\BibitemShut {NoStop}%
\bibitem [{\citenamefont {Aseev}\ \emph {et~al.}(2019)\citenamefont {Aseev},
  \citenamefont {Marra}, \citenamefont {Stano}, \citenamefont {Klinovaja},\
  and\ \citenamefont {Loss}}]{Aseev2019}%
  \BibitemOpen
  \bibfield  {author} {\bibinfo {author} {\bibfnamefont {P.~P.}\ \bibnamefont
  {Aseev}}, \bibinfo {author} {\bibfnamefont {P.}~\bibnamefont {Marra}},
  \bibinfo {author} {\bibfnamefont {P.}~\bibnamefont {Stano}}, \bibinfo
  {author} {\bibfnamefont {J.}~\bibnamefont {Klinovaja}}, \ and\ \bibinfo
  {author} {\bibfnamefont {D.}~\bibnamefont {Loss}},\ }\href@noop {} {\bibfield
   {journal} {\bibinfo  {journal} {Phys. Rev. B}\ }\textbf {\bibinfo {volume}
  {99}},\ \bibinfo {pages} {205435} (\bibinfo {year} {2019})}\BibitemShut
  {NoStop}%
\bibitem [{\citenamefont {Dom{\'i}nguez}\ \emph {et~al.}(2012)\citenamefont
  {Dom{\'i}nguez}, \citenamefont {Hassler},\ and\ \citenamefont
  {Platero}}]{Dominguez2012}%
  \BibitemOpen
  \bibfield  {author} {\bibinfo {author} {\bibfnamefont {F.}~\bibnamefont
  {Dom{\'i}nguez}}, \bibinfo {author} {\bibfnamefont {F.}~\bibnamefont
  {Hassler}}, \ and\ \bibinfo {author} {\bibfnamefont {G.}~\bibnamefont
  {Platero}},\ }\href@noop {} {\bibfield  {journal} {\bibinfo  {journal} {Phys.
  Rev. B.}\ }\textbf {\bibinfo {volume} {86}},\ \bibinfo {pages} {140503(R)}
  (\bibinfo {year} {2012})}\BibitemShut {NoStop}%
\bibitem [{\citenamefont {Pikulin}\ and\ \citenamefont
  {Nazarov}(2012)}]{Pikulin2012}%
  \BibitemOpen
  \bibfield  {author} {\bibinfo {author} {\bibfnamefont {D.~I.}\ \bibnamefont
  {Pikulin}}\ and\ \bibinfo {author} {\bibfnamefont {Y.~V.}\ \bibnamefont
  {Nazarov}},\ }\href@noop {} {\bibfield  {journal} {\bibinfo  {journal} {Phys.
  Rev. B.}\ }\textbf {\bibinfo {volume} {86}},\ \bibinfo {pages} {140504(R)}
  (\bibinfo {year} {2012})}\BibitemShut {NoStop}%
\end{thebibliography}%


\begin{thebibliography}{9}%
\makeatletter
\providecommand \@ifxundefined [1]{%
 \@ifx{#1\undefined}
}%
\providecommand \@ifnum [1]{%
 \ifnum #1\expandafter \@firstoftwo
 \else \expandafter \@secondoftwo
 \fi
}%
\providecommand \@ifx [1]{%
 \ifx #1\expandafter \@firstoftwo
 \else \expandafter \@secondoftwo
 \fi
}%
\providecommand \natexlab [1]{#1}%
\providecommand \enquote  [1]{``#1''}%
\providecommand \bibnamefont  [1]{#1}%
\providecommand \bibfnamefont [1]{#1}%
\providecommand \citenamefont [1]{#1}%
\providecommand \href@noop [0]{\@secondoftwo}%
\providecommand \href [0]{\begingroup \@sanitize@url \@href}%
\providecommand \@href[1]{\@@startlink{#1}\@@href}%
\providecommand \@@href[1]{\endgroup#1\@@endlink}%
\providecommand \@sanitize@url [0]{\catcode `\\12\catcode `\$12\catcode
  `\&12\catcode `\#12\catcode `\^12\catcode `\_12\catcode `\%12\relax}%
\providecommand \@@startlink[1]{}%
\providecommand \@@endlink[0]{}%
\providecommand \url  [0]{\begingroup\@sanitize@url \@url }%
\providecommand \@url [1]{\endgroup\@href {#1}{\urlprefix }}%
\providecommand \urlprefix  [0]{URL }%
\providecommand \Eprint [0]{\href }%
\providecommand \doibase [0]{http://dx.doi.org/}%
\providecommand \selectlanguage [0]{\@gobble}%
\providecommand \bibinfo  [0]{\@secondoftwo}%
\providecommand \bibfield  [0]{\@secondoftwo}%
\providecommand \translation [1]{[#1]}%
\providecommand \BibitemOpen [0]{}%
\providecommand \bibitemStop [0]{}%
\providecommand \bibitemNoStop [0]{.\EOS\space}%
\providecommand \EOS [0]{\spacefactor3000\relax}%
\providecommand \BibitemShut  [1]{\csname bibitem#1\endcsname}%
\let\auto@bib@innerbib\@empty
%</preamble>
\bibitem [{\citenamefont {Klinovaja}\ and\ \citenamefont
  {Loss}(2012)}]{Klinovaja2012}%
  \BibitemOpen
  \bibfield  {author} {\bibinfo {author} {\bibfnamefont {J.}~\bibnamefont
  {Klinovaja}}\ and\ \bibinfo {author} {\bibfnamefont {D.}~\bibnamefont
  {Loss}},\ }\href@noop {} {\bibfield  {journal} {\bibinfo  {journal} {Phys.
  Rev. B}\ }\textbf {\bibinfo {volume} {86}},\ \bibinfo {pages} {085408}
  (\bibinfo {year} {2012})}\BibitemShut {NoStop}%
\bibitem [{\citenamefont {San-Jose}\ \emph {et~al.}(2013)\citenamefont
  {San-Jose}, \citenamefont {Cayao}, \citenamefont {Prada},\ and\ \citenamefont
  {Aguado}}]{Aguado2013}%
  \BibitemOpen
  \bibfield  {author} {\bibinfo {author} {\bibfnamefont {P.}~\bibnamefont
  {San-Jose}}, \bibinfo {author} {\bibfnamefont {J.}~\bibnamefont {Cayao}},
  \bibinfo {author} {\bibfnamefont {E.}~\bibnamefont {Prada}}, \ and\ \bibinfo
  {author} {\bibfnamefont {R.}~\bibnamefont {Aguado}},\ }\href@noop {}
  {\bibfield  {journal} {\bibinfo  {journal} {New J. Phys.}\ }\textbf {\bibinfo
  {volume} {15}},\ \bibinfo {pages} {075019} (\bibinfo {year}
  {2013})}\BibitemShut {NoStop}%
\bibitem [{\citenamefont {Murthy}\ \emph {et~al.}(2020)\citenamefont {Murthy},
  \citenamefont {Kurilovich}, \citenamefont {van Heck}, \citenamefont
  {Glazman},\ and\ \citenamefont {Nayak}}]{Glazman2020}%
  \BibitemOpen
  \bibfield  {author} {\bibinfo {author} {\bibfnamefont {C.}~\bibnamefont
  {Murthy}}, \bibinfo {author} {\bibfnamefont {P.~D.}\ \bibnamefont
  {Kurilovich}, \bibfnamefont {V.~D. and.~Kurilovich}}, \bibinfo {author}
  {\bibfnamefont {B.}~\bibnamefont {van Heck}}, \bibinfo {author}
  {\bibfnamefont {L.~I.}\ \bibnamefont {Glazman}}, \ and\ \bibinfo {author}
  {\bibfnamefont {C.}~\bibnamefont {Nayak}},\ }\href@noop {} {\bibfield
  {journal} {\bibinfo  {journal} {Phys. Rev. B.}\ }\textbf {\bibinfo {volume}
  {101}},\ \bibinfo {pages} {224501} (\bibinfo {year} {2020})}\BibitemShut
  {NoStop}%
\bibitem [{\citenamefont {Fu}\ and\ \citenamefont {Kane}(2009)}]{Kane2009}%
  \BibitemOpen
  \bibfield  {author} {\bibinfo {author} {\bibfnamefont {L.}~\bibnamefont
  {Fu}}\ and\ \bibinfo {author} {\bibfnamefont {C.~L.}\ \bibnamefont {Kane}},\
  }\href@noop {} {\bibfield  {journal} {\bibinfo  {journal} {Phys. Rev. B}\
  }\textbf {\bibinfo {volume} {79}},\ \bibinfo {pages} {161408} (\bibinfo
  {year} {2009})}\BibitemShut {NoStop}%
\bibitem [{\citenamefont {Cayao}\ \emph {et~al.}(2017)\citenamefont {Cayao},
  \citenamefont {San-Jose}, \citenamefont {Black-Schaffer}, \citenamefont
  {Aguado},\ and\ \citenamefont {Prada}}]{Prada2017}%
  \BibitemOpen
  \bibfield  {author} {\bibinfo {author} {\bibfnamefont {J.}~\bibnamefont
  {Cayao}}, \bibinfo {author} {\bibfnamefont {P.}~\bibnamefont {San-Jose}},
  \bibinfo {author} {\bibfnamefont {A.~M.}\ \bibnamefont {Black-Schaffer}},
  \bibinfo {author} {\bibfnamefont {R.}~\bibnamefont {Aguado}}, \ and\ \bibinfo
  {author} {\bibfnamefont {E.}~\bibnamefont {Prada}},\ }\href@noop {}
  {\bibfield  {journal} {\bibinfo  {journal} {Phys. Rev. B}\ }\textbf {\bibinfo
  {volume} {96}},\ \bibinfo {pages} {205425} (\bibinfo {year}
  {2017})}\BibitemShut {NoStop}%
\bibitem [{\citenamefont {Pikulin}\ and\ \citenamefont
  {Nazarov}(2012)}]{Pikulin2012}%
  \BibitemOpen
  \bibfield  {author} {\bibinfo {author} {\bibfnamefont {D.~I.}\ \bibnamefont
  {Pikulin}}\ and\ \bibinfo {author} {\bibfnamefont {Y.~V.}\ \bibnamefont
  {Nazarov}},\ }\href@noop {} {\bibfield  {journal} {\bibinfo  {journal} {Phys.
  Rev. B.}\ }\textbf {\bibinfo {volume} {86}},\ \bibinfo {pages} {140504(R)}
  (\bibinfo {year} {2012})}\BibitemShut {NoStop}%
\bibitem [{\citenamefont {Dom{\'i}nguez}\ \emph {et~al.}(2012)\citenamefont
  {Dom{\'i}nguez}, \citenamefont {Hassler},\ and\ \citenamefont
  {Platero}}]{Dominguez2012}%
  \BibitemOpen
  \bibfield  {author} {\bibinfo {author} {\bibfnamefont {F.}~\bibnamefont
  {Dom{\'i}nguez}}, \bibinfo {author} {\bibfnamefont {F.}~\bibnamefont
  {Hassler}}, \ and\ \bibinfo {author} {\bibfnamefont {G.}~\bibnamefont
  {Platero}},\ }\href@noop {} {\bibfield  {journal} {\bibinfo  {journal} {Phys.
  Rev. B.}\ }\textbf {\bibinfo {volume} {86}},\ \bibinfo {pages} {140503(R)}
  (\bibinfo {year} {2012})}\BibitemShut {NoStop}%
\bibitem [{\citenamefont {Levenson-Falk}\ \emph {et~al.}(2014)\citenamefont
  {Levenson-Falk}, \citenamefont {Kos}, \citenamefont {Vijay}, \citenamefont
  {Glazman},\ and\ \citenamefont {Siddiqi}}]{Falk2014}%
  \BibitemOpen
  \bibfield  {author} {\bibinfo {author} {\bibfnamefont {E.~M.}\ \bibnamefont
  {Levenson-Falk}}, \bibinfo {author} {\bibfnamefont {F.}~\bibnamefont {Kos}},
  \bibinfo {author} {\bibfnamefont {R.}~\bibnamefont {Vijay}}, \bibinfo
  {author} {\bibfnamefont {L.}~\bibnamefont {Glazman}}, \ and\ \bibinfo
  {author} {\bibfnamefont {I.}~\bibnamefont {Siddiqi}},\ }\href@noop {}
  {\bibfield  {journal} {\bibinfo  {journal} {Phys. Rev. Lett.}\ }\textbf
  {\bibinfo {volume} {112}},\ \bibinfo {pages} {047002} (\bibinfo {year}
  {2014})}\BibitemShut {NoStop}%
\bibitem [{\citenamefont {Tinkham}(1996)}]{Tinkham1996}%
  \BibitemOpen
  \bibfield  {author} {\bibinfo {author} {\bibfnamefont {M.}~\bibnamefont
  {Tinkham}},\ }in\ \href@noop {} {\emph {\bibinfo {booktitle} {Introduction to
  Superconductivity}}}\ (\bibinfo  {publisher} {Second Edition, McGraw-Hill
  Book Co.},\ \bibinfo {year} {1996})\BibitemShut {NoStop}%
\end{thebibliography}%

\end{document}

\title{Supplemental Material:\\
``Quasiparticle poisoning in trivial and topological Josephson junctions"}
\author{Aleksandr E. Svetogorov}
\author{Daniel Loss}
\author{Jelena Klinovaja}
\affiliation{Department of Physics, University of Basel, Klingelbergstrasse, 4056 Basel, Switzerland}
\maketitle
\section{The low-energy spectrum of a Josephson junction based on a proximitized semiconducting nanowire with strong SOI close to the phase transition}\label{app:abs}

In this section we provide a detailed analysis of the low-energy spectrum of a topological Josephson junction (JJ). We analyze the scaling of the phase potential amplitude ($E_J$ for trivial phase and $E_M$ for topological) with the normal state transmission amplitude $D_N$ to validate our analysis of phase evolution after the parity switch in the main text. Our starting point is a linearization procedure~\cite{Klinovaja2012} with subsequent calculation of the low-energy spectrum~\cite{Aguado2013,Glazman2020} for a JJ based on a semiconducting nanowire with strong Rashba spin-orbit interaction (SOI) of strength $\alpha$, as in this regime the junction is in the Andreev limit ($\Delta\ll E_F\sim m\alpha^2$, where $E_F$ is the Fermi energy measured from the bottom of conduction band and is set by the characteristic energy of SOI $\sim m\alpha^2$, where $m$ is the effective electron band mass). Two further ingredients are the Zeeman field $V_Z\ll m\alpha^2$ and the proximity gap $\Delta$. An additional assumption we make to simplify the analysis is that the Fermi level is exactly in the middle of the Zeeman gap. For the case of a short junction region, modelled as a delta-function barrier, one can write down the equation determining the bound states as~\cite{Glazman2020}: 
\begin{equation}
\Lambda(E,\phi,D_N,V_Z)=0,
\end{equation}
where  $\Lambda $ is given by
\begin{align}
&\Lambda(E,\phi,D_N,V_Z)=\Lambda_0(E,\phi,V_Z)+(1-D_N)\Lambda_1(E,\phi,V_Z)+(1-D_N)^2\Lambda_2(E,\phi,V_Z)+D_N\Lambda_\gamma(E,\phi,V_Z) \sin^2\gamma;\\
&\Lambda_0(E,\phi,V_Z)=\left[\Delta^{2}\cos^{2}\frac{\phi}{2}-E^{2}\right]\left[\sqrt{\Delta_{-}^{2}-E^{2}}\sqrt{\Delta_{+}^{2}-E^{2}}\left(1+\cos^{2}\frac{\phi}{2}\right)-\left(E^{2}+\Delta_{-}\Delta_{+}\right)\sin^{2}\frac{\phi}{2}\right],\\
&\Lambda_1(E,\phi,V_Z)=\left[\left(E^{2}+\Delta_{-}\Delta_{+}\right)\left(2\Delta^{2}\cos^{2}\frac{\phi}{2}-E^{2}\right)-2\Delta^{2}E^{2}\right]\sin^{2}\frac{\phi}{2}\nonumber\\
&+\sqrt{\Delta^{2}-E^{2}}\sqrt{\Delta_{-}^{2}-E^{2}}\left[\Delta\Delta_{+}\cos^{2}\frac{\phi}{2}-E^{2}\right]
+\sqrt{\Delta^{2}-E^{2}}\sqrt{\Delta_{+}^{2}-E^{2}}\left[\Delta\Delta_{-}\cos^{2}\frac{\phi}{2}-E^{2}\right]\nonumber\\
&-\sqrt{\Delta_{-}^{2}-E^{2}}\sqrt{\Delta_{+}^{2}-E^{2}}\left[2\Delta^{2}\cos^{4}\frac{\phi}{2}-E^{2}\left(1+\cos^{2}\frac{\phi}{2}\right)\right],\\
&\Lambda_2(E,\phi,B)=\Delta^{2}\left(E^{2}+\Delta_{-}\Delta_{+}\right)\sin^{4}\frac{\phi}{2}+\frac{1}{2}\left(\Delta_{-}\Delta_{+}-E^{2}\right)\left[\Delta^{2}\cos\phi-E^{2}\right]\nonumber\\
&-\frac{1}{2}B^{2}E^{2}-\frac{1}{2}\sqrt{\Delta^{2}-E^{2}}\sqrt{\Delta_{-}^{2}-E^{2}}\left[\Delta\Delta_{+}\cos\phi-E^{2}\right]
-\frac{1}{2}\sqrt{\Delta^{2}-E^{2}}\sqrt{\Delta_{+}^{2}-E^{2}}\left[\Delta\Delta_{-}\cos\phi-E^{2}\right]\nonumber\\
&+\frac{1}{2}\sqrt{\Delta_{-}^{2}-E^{2}}\sqrt{\Delta_{+}^{2}-E^{2}}\left[\Delta^{2}\left(1-\frac{1}{2}\sin^{2}\phi\right)-E^{2}\right],\\
&\Lambda_\gamma(E,\phi,V_Z)=\left[\Delta^{2}\cos^{2}\frac{\phi}{2}-E^{2}\right]\left[-\sqrt{\Delta_{-}^{2}-E^{2}}\sqrt{\Delta_{+}^{2}-E^{2}}+\Delta_{-}\Delta_{+}-E^{2}\right].
\end{align}
\begin{figure}[h]
\includegraphics[scale=0.72]{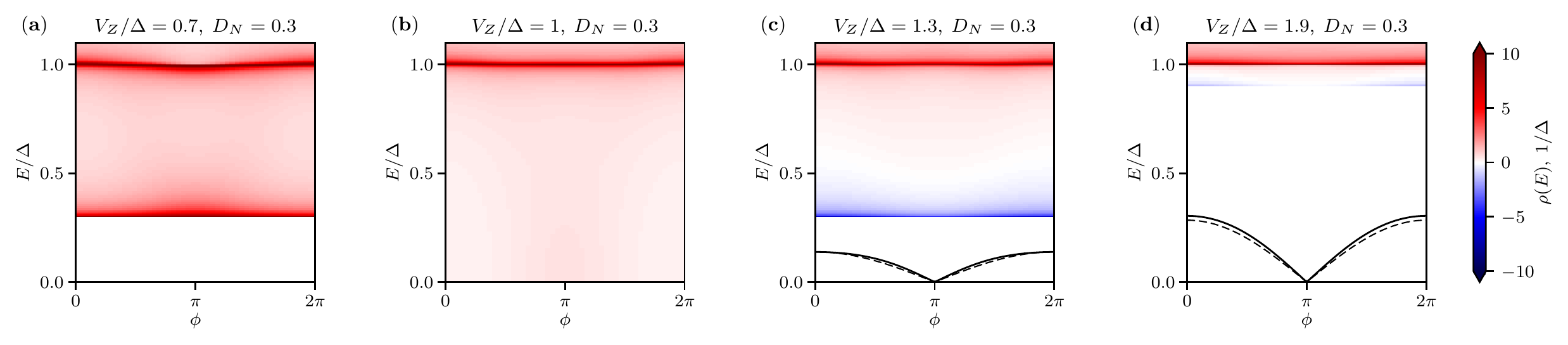}
\includegraphics[scale=0.67]{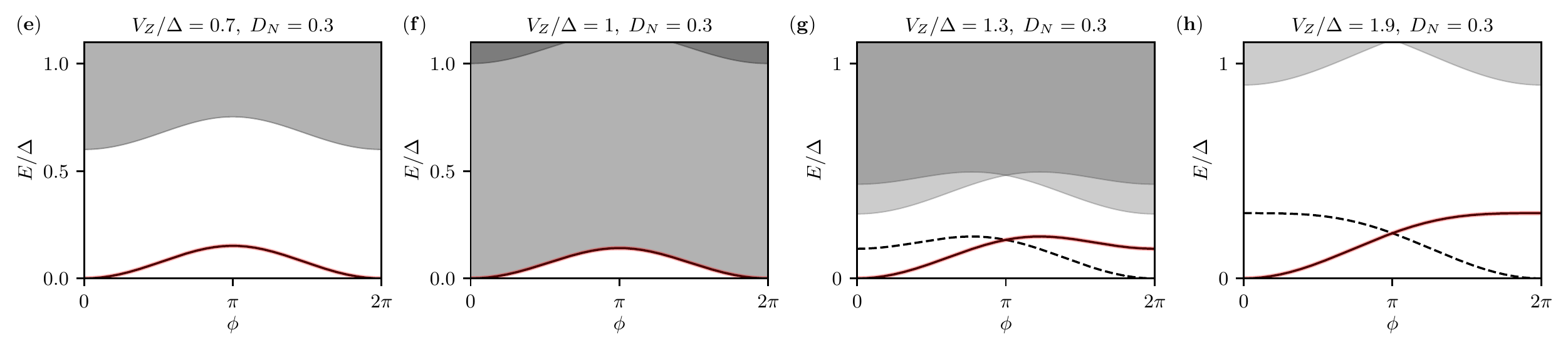}
\caption{Upper panels (a-d) show the single-particle  spectrum at different values of Zeeman field $V_Z$ as function of phase $\phi$, corresponding to (a) trivial phase, (b) transition,  and  (c, d) topological phase. The colour over $|V_Z-\Delta|$ indicates the local density of states, while the black line below the gap corresponds to the bound state. In the topological phase (c, d) the dashed line corresponds to our analytical solution; we note that both parity solutions are depicted, one before $\pi$ another one after $\pi$. In the trivial phase (a) the bound state is absent for small $D_N=0.3$. The lower panel (e-h) shows the many-body spectrum of the junction vs. $\phi$ at different values of $V_Z$; the solid red line represents the ground state, the grey area is the continuum, and the dotted line represents an odd-parity state. In the trivial state (e) at $V_Z/\Delta=0.7$ the spectrum consists of the ground state and no discrete excited states up to the continuum (since there are no bound states); panel (f) shows the gapless spectrum at the phase transition; panel (g) corresponds to the topological phase at $V_Z/\Delta=1.3$, each second local minimum of the ground state is suppressed by a $4\pi$-periodic contribution, and panel (h) corresponds to the topological phase at $V_Z/\Delta=1.9$, where only $4\pi$-periodic local minima remain.}\label{fig:spectrum}
\end{figure}
Here $\Delta_-=|\Delta-V_Z|$, the gap of the system, and $\Delta_+=\Delta+V_Z$ are introduced for simplicity, $\gamma$ is the forward scattering phase, for the toy model of a point-like potential it is given by \begin{equation}\label{eq:gamma}
\gamma=-\arctan\sqrt{\frac{1-D_N}{D_N}}.
\end{equation}

In this system the continuum consisting of the scattering states above $|\Delta-V_Z|$ (or more generally the smallest gap in the system $\mathrm{min}\left[|\Delta-V_Z|,\Delta\right]$) should be taken into account, as it possesses a phase dependence (one can see it as a bound state merged into the continuum) and, therefore, affects the spectrum of the system already starting from the ground state. This phase-dependent contribution to the density of states in the continuum can be calculated via the same $\Lambda$ expression~\cite{Glazman2020}:
\begin{equation}
\delta\rho(E,\phi)=\frac{1}{2\pi{i}}\frac{\partial}{\partial{E}}\frac{\Lambda^*(E,\phi,D_N,V_Z)}{\Lambda(E,\phi,D_N,V_Z)}.
\end{equation}
The spectrum in the high-transparency limit was analyzed in~\cite{Aguado2013,Glazman2020}.

\begin{figure}
\includegraphics[width=8cm]{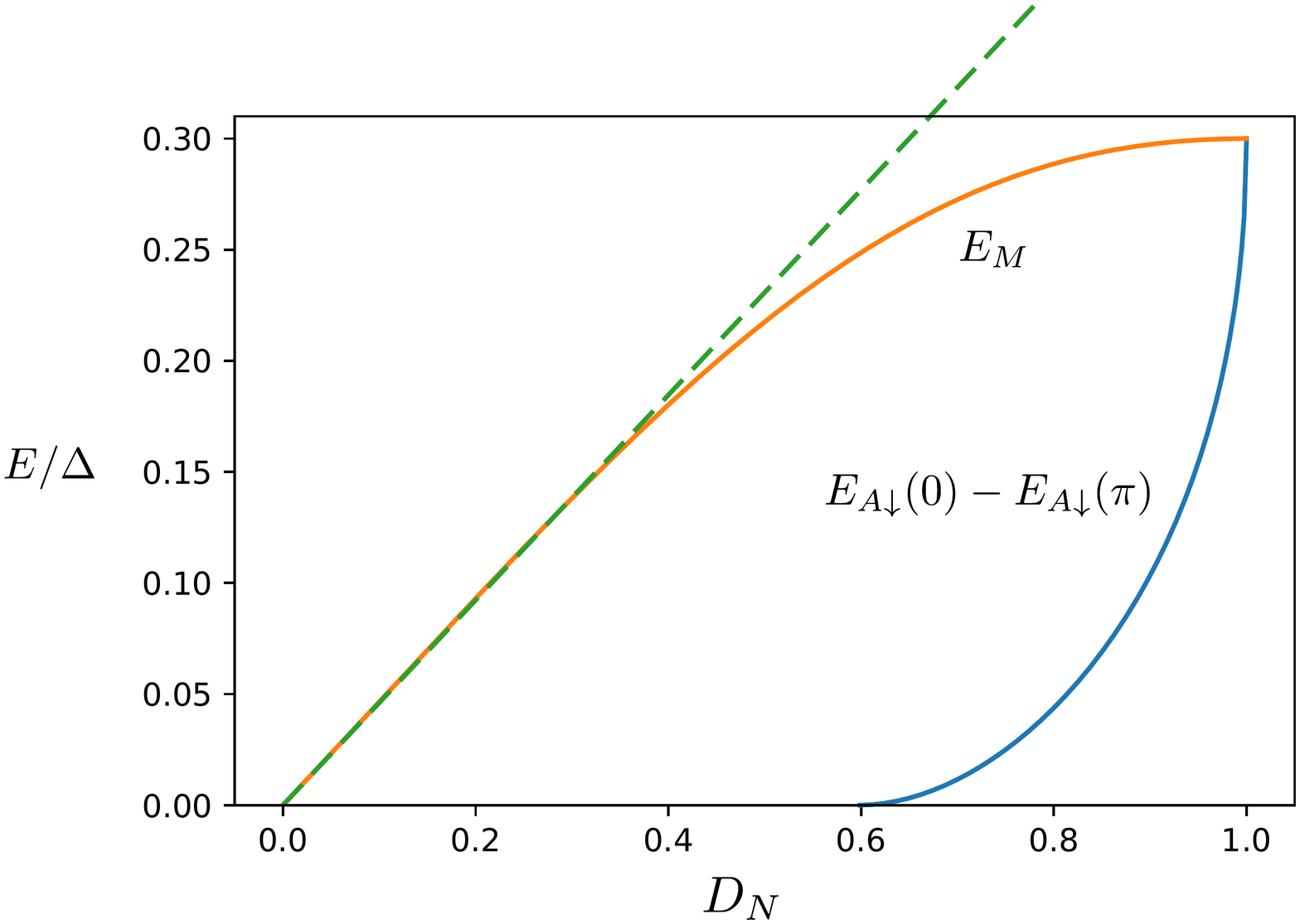}
\caption{Dependence of the bound state amplitude on transmission amplitude $D_N$. The orange branch corresponds to the topological phase $V_Z=1.3\Delta$, green dashed line is the analytical solution for low transparency;  blue curve represents lowest ABS energy amplitude in the trivial phase for $V_Z=0.7\Delta$ (difference of ABS energy at $\phi=0$, when it touches the gap, and at $\phi=\pi$, when it is minimal). One can see that there is no ABS for low $D_N$, then starting from some value of transmission the bound state amplitude grows up to the bulk gap size $\Delta_c=\Delta-V_Z$.}\label{fig:ABS_scaling}
\end{figure}

Here, we study the low-transparency limit, as it allows us to perform some estimations due to the simple cosine dependences on the phase (tunneling regime). To begin with, we  expand
\begin{equation}
\sin^{2}\gamma=1-\frac{D_N}{4}+O(D_N^2).
\end{equation}
Searching for the zeroes of $\Lambda(E,\phi,D_N,V_Z)$ numerically, we see that the lowest ABS sticks to the gap, as expected, while the second ABS is completely merged with the continuum. The remaining ABS does not scale linearly with $D_N$,  as in a trivial spin-degenerate ABS, but rapidly decays to zero with increase of $V_Z$. For each transmission amplitude $D_N$ there is a value $V_Z^{bs}$, starting from which there is no bound state in the trivial phase,  this value can be found as a solution of the equation: 
\begin{equation}
\Lambda(\Delta-V_Z^{bs},\pi,D_N,V_Z^{bs})=0.
\end{equation}
From the numerical analysis we can conclude that in the low-transparency case, the system has no bound states in a large vicinity of the phase transition, i.e. for $D_N=0.3$ the bound state disappears already at $V_Z=0.4328 \Delta$. Therefore, we state that in a large vicinity of the transition all the phase dependences of the low-energy spectrum are determined by the phase-dependent density of states of the continuum, which allows us to state (in the main text) that there are no effects due to quasiparticle poisoning around the phase transition. Moreover, what is also interesting is that the total contribution (ABSs plus continuum) to the ground state in the trivial phase has a weak dependence on the magnetic field, which can be seen as if the continuum is acquiring a phase-dependent contribution from the ABSs merged into it. On the contrary, in the topological phase  the only bound state amplitude scales as $D_N$ at low transparency, as the continuum is gapped away from the bound state. We have calculated the amplitude in the low transparency limit expanding $\Lambda(E,\phi,D_N,V_Z)$ analytically in $D_N$ and $E/(V_Z-\Delta)$ (see Fig.~\ref{fig:ABS_scaling}):
\begin{equation}
E\approx2\frac{D_N\Delta}{V_Z}(V_Z-\Delta)\cos\frac\phi{2},
\end{equation}
which is Eq.~(4) in the main text. What is a bit counterintuitive at first sight is the linear scaling with $D_N$, as naively it is expected to be the same as in Ref.~\cite{Kane2009} $\sim\sqrt{D_N}$, which corresponds to the single electron tunneling amplitude. The reason for this is that in Ref.~\cite{Kane2009} the edge states are supposed to carry only two helical modes, which is valid only at large Zeeman fields $V_Z\gg\Delta$~\cite{Glazman2020}, where our formula is not applicable, as the SOI energy is no longer the largest energy scale in the system. 

Now we can write down the many-body spectrum of the system. For the low-energy states it is described by the occupation of bound states and continuum states below the superconducting gap $\Delta$:
\begin{equation}
H(\phi)=E_b(\phi)\left(n_b-\frac12\right)+\sum_iE_i(\phi)\left(n_i-\frac12\right)+H_0,
\end{equation}
where $E_b(\phi)$ is the bound state energy - either the lowest ABS in the trivial phase, or the bound state formed by two MBSs in the topological phase - $n_b$ is the occupation of the state; $|\Delta-V_Z|<E_i(\phi)<\Delta$ are the energies of the continuum states, $n_i$ is the number operator for fermions in each continuum state; $H_0$ is the phase-independent contribution. From the numerical solution we deduce that a parity switch close to the phase transition can only slightly modify the phase dependence of the system energy, as $E_b$ is suppressed for small $D_N$ or absent completely (all ABSs merging with continuum) and the main contribution to the low-energy spectrum comes from the scattering states in the continuum between the system gap $\Delta-V_Z$ and superconducting gap $\Delta$. On the other side, deep inside the topological phase, a $4\pi$-periodic bound state contribution dominates over the continuum contribution and, therefore, each parity switching event significantly changes the phase dependence of the system energy. The full picture of the spectrum evolution with Zeeman field can be seen in Fig.~\ref{fig:spectrum}.

In a junction based on proximitized semiconducting nanowires with weak SOI, the Andreev limit is violated. Therefore, the regime is very hard to analyze analytically, however, we expect the qualitative behaviour of the system to remain the same. At zero Zeeman field two degenerate ABSs are present, the significant difference with strong SOI regime is the absence of zero-energy crossings at $\phi=\pi$ even for perfect transparency of the junction $D_N=1$. In the presence of an external magnetic field, the ABSs split and get different amplitudes in the phase, as well as a phase-dependent contributions from the continuum appear. Close to the topological phase transition the continuum contribution to the ground state is dominating (on both sides of the transition). In the topological phase one would get $4\pi$-periodic term due to hybridization of the two MBSs on the junction, however, the amplitude is determined not by the superconducting gap and Zeeman field $\Delta_c=V_Z-\Delta$, as in the strong SOI limit, but also by the SOI energy scale $E_{SO}=m\alpha^2/2$~\cite{Klinovaja2012,Prada2017}:
\begin{equation}
\Delta_c\approx2\sqrt{\frac{E_{SO}}{V_Z}}\Delta.
\end{equation}
Nevertheless, deep inside the topological phase the MBS contribution is again dominating in the phase potential, therefore, the spectrum has the same features which are crucial for the observation of the effects discussed in the main text.

\section{Finite-size effects in the topological phase}
 \begin{figure}[h]
\includegraphics[width=12cm]{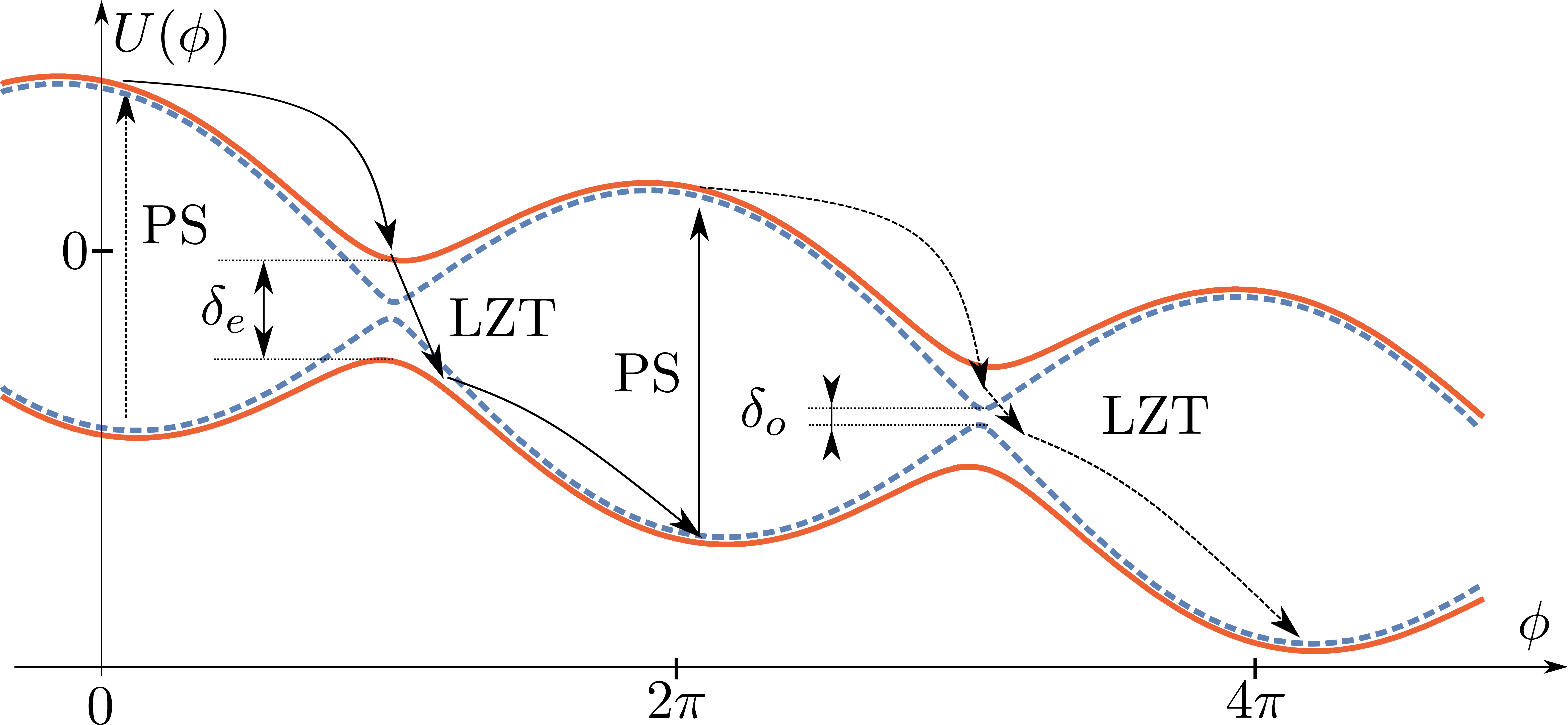}
\caption{The phase potential $U$ for two different parities (solid red line for even parity and dashed blue for odd) as function of phase $\phi$. Here we have chosen smaller splittings in the odd state than in the even state $\delta_o<\delta_e$, however, this does not need to be the case in general. Arrows correspond to phase evolution: vertical arrows marked with PS correspond to parity switches, arrows along the phase potential branches represent phase relaxation, LZT represents Landau-Zener transitions at the anticrossings. All the solid arrows correspond to phase evolution in the even state (or starting in even state for phase switch), dashed arrows to that of the odd state. }\label{fig:pot_M}
\end{figure} 
Here we analyze finite-size effects in the setup discussed in the main text and show that they do not significantly affect the proposed measurement. If there is a finite overlap with the MBSs on the outer edges of the topological part of the system, crossings at $\phi\approx(2k+1)\pi$ turn into anticrossings~\cite{Pikulin2012,Dominguez2012}. Strictly speaking, there are now four energy levels instead of two:
\begin{equation}
E^e_\pm=\pm\frac{1}{2}\sqrt{E_M^2\cos^2\frac{\phi}{2}+(\delta_L+\delta_R)^2},
\end{equation}
\begin{equation}
E^o_\pm=\pm\frac{1}{2}\sqrt{E_M^2\cos^2\frac{\phi}{2}+(\delta_L-\delta_R)^2},
\end{equation}
where the superscript $e$ ($o$) stands for even (odd) parity solution, $\delta_L$ ($\delta_R$) is the coupling with the MBS on the left (right) edge of the topological nanowire. What is important now is that the parity is the parity of the superposition of all four MBSs. Another issue worth mentioning is that $\delta_R$ and $\delta_L$ depend on the exact geometry of the nanowires forming the junction and can be both positive or negative. However, for each parity (which is fixed between parity switching events) the spectrum consist of two branches with fixed anticrossing $\delta=\left|\delta_L\pm\delta_R\right|$ (the sign of it being defined by the parity). If the phase evolution (relaxation) is fast enough in comparison to the splitting $\delta$ at the anticrossing, $\tau_{2\pi}^{-1}\gg\delta^2/(2E_M)$, a Landau-Zener transition (LZT) would let the system reach the lower branch and relax to a new $2\pi$-shifted minimum as in the perfectly $4\pi$-periodic case. However, if $R/R_Q\ll\delta^2/E_M^2$, which can be the case for strongly overdamped junctions and for finite $\delta$, no LZT occurs and the state stays in the upper branch, thereby relaxing to  the minimum at the anticrossing, giving rise to a voltage pulse of size $V=\frac{1}{2e}\frac{\pi}{\tau_{\pi}}$, where $\tau_\pi$ is the characteristic time for relaxation to a new $\pi$-shifted minimum. The next quasiparticle poisoning event will transfer the state to the lower branch with subsequent relaxation to a minimum, which also results in a $\pi$ phase shift. Therefore, the average voltage will be twice smaller than in the $4\pi$-periodic case [cf. Eq. (6) in the main text]:
\begin{equation}\label{eq:voltage}
\langle{V}\rangle\approx\frac{1}{2e}\frac{\pi}{\tau_{qp}^{top}}.
\end{equation}
If one can fine tune the overlap of the MBSs so that in one parity state the splitting $\delta$ is large enough to suppress LZT and in another parity state it is very small (i.e. $\delta_L\approx\delta_R$ and $R/R_Q\ll4\delta_L^2/E_M^2$), then it is possible to achieve an intermediate regime, where the average voltage is
\begin{equation}
\langle{V}\rangle\approx\frac{1}{2e}\frac{3\pi}{2\tau_{qp}^{top}},
\end{equation}
as after the parity switch the phase will relax by $\pi$ or $2\pi$ depending on the parity.

\section{Voltage Pulses in the trivial phase}

In this section we provide a detailed description of the quasiparticle poisoning effect in the trivial phase to supplement the discussion of the effect in the main text. If the junction is in the trivial phase, quasiparticles can be trapped in ABSs or recombine pairwise into the condensate of Cooper pairs. Both processes are rather slow (in comparison to getting trapped in the topological phase). Recombination is slow, since the recombination rate depends on the quasiparticle density quadratically ($\sim x^2_{qp}$, where $x_{qp}=n_{qp}/n_{cp}\ll1$ is the ratio of quasiparticle density and Cooper pair density), while the trapping rate to an ABS is low due to the small energy gap between the ABS energy and the continuum edge. If the bias current is zero, $I=0$, the bound states touch the continuum edge at the phase values corresponding to the minima of the phase potential $\phi_{min}=2\pi k$ ($k$ is integer), therefore, a quasiparticle cannot be trapped.  For  small bias current, $I\ll I_c$, the minima of the phase potential are slightly shifted, i.e. $\phi_{min}\approx 2\pi k+\frac{I}{2eE_J}$, which results in a small energy gap between the continuum edge and the bound state $\epsilon=I^2/(4e^2E_J)\ll\Delta$ at $\phi_{min}$, which allows quasiparticle trapping in the bound states. As a result, at zero magnetic field, when a quasiparticle is trapped in an ABS, the only conducting channel will get poisoned, which would result in a running (resistive) state with voltage $V=IR$. Subsequently the system will relax back to the ground state due to recombination of the trapped quasiparticle with a quasiparticle from the continuum on a time scale $\tau_r\ll\tau_{qp}^{triv}$. What is essential here is that the recombination rate of a quasiparticle trapped in the bound state with a quasiparticle from the continuum is linear in the quasiparticle density $\sim x_{qp}$, as only one quasiparicle from the continuum is required. Therefore, the rate of such a recombination is sufficiently larger than the recombination rate for quasiparticle pairs from the continuum. At the same time, this recombination process releases the energy $\sim\Delta$ (in the form of a phonon), which makes it significantly more probable than the trapping of another quasiparticle from the continuum into the second ABS (releasing a phonon with much smaller energy $\epsilon\ll\Delta$), as normally at low energies the phonon density-of-states is increasing with energy (usually for metals it is quadratic $\nu_{ph}\sim E^2$). This can be seen from a simple estimate of the rate for a quasiparticle getting trapped into an ABS with energy $E_A$ using a simplified model of a bulk superconductor~\cite{Falk2014}:
\[
\Gamma_t(\epsilon)=\pi x_{qp}\Delta\alpha_{e-ph}^2\nu_{ph}(\Delta-E_A),
\]
where $\alpha_{e-ph}$ is the electron-phonon interaction matrix element. For recombination processes, when one quasiparticle is trapped to an ABS and another one is coming from the continuum edge, one can use the same formula for the recombination rate $\Gamma_r$, except replacing the phonon energy $\Delta-E_A=\epsilon$ by $\Delta+E_A\approx2\Delta\gg\epsilon$. Under the assumption of a quadratic dependence of the phonon density-of-states on energy one gets the ratio:
\begin{equation}
\Gamma_r/\Gamma_t^{triv}\approx\frac{(2\Delta)^2}{\epsilon^2}\gg1.
\end{equation}
A similar formula can be used to estimate the ratio between quasiparticle trapping rates in the trivial and topological phases, which corresponds to Eq.~(7) in the main text:
\begin{equation}
\Gamma_t^{top}/\Gamma_t^{triv}\approx\frac{(\mathrm{min}\left[V_Z-\Delta,\Delta\right]-E_M)^2}{\epsilon^2}\approx\frac{(2e)^4E_J^2(1-D_N)^2\Delta^2}{I^4}\gg1,
\end{equation}
since deep in the topological phase ($V_Z>2\Delta$) the bound state is always separated from the continuum (except for the case of perfect transparency), while the energy gap between the ABS and the continuum in the trivial phase is small: $\epsilon\approx I^2/(4e^2E_J)\ll \Delta$. We should note that quasiparticle poisoning rates in the topological phase, $\Gamma_t^{top}$, and relaxation rates in the trivial phase, $\Gamma_r$, (due to recombination of the trapped quasiparticle) can be of the same order, since the released phonons carry away energy of the order of the gap in both cases.

Turning on a magnetic field lowers the continuum edge ($\Delta-V_Z$) as well as split the ABSs. As a result, with one quasiparticle being trapped in one of the ABSs, the phase potential is no longer linear. According to numerical results provided in Ref.~\cite{Aguado2013,Glazman2020} and the beginning of the supplemental material, for low fields a good approximation is to represent the energetically higher
ABS plus continuum just as an effective higher ABS (which is merged
in the continuum between $\Delta-B$ and $\Delta)$. Then, we can represent
the lower ABS energy as $E_{A2}=\left(\Delta-V_Z\right)-E_{J2}(1-\cos\phi)$,
while the combination of higher ABS and continuum contribution is given by
$E_{A1}=\Delta-E_{J1}(1-\cos\phi)$. Moreover, in the Andreev limit we can express the
amplitudes through the transmission coefficient $D_N$ (as in general $E_{A}=\Delta\sqrt{1-D_N\sin^{2}\frac{\phi}{2}}\approx\Delta-\frac{\Delta D_N}{4}\left(1-\cos\phi\right)$).
Then, $E_{J2}\approx D_N\left(\Delta-V_Z\right)/4$, $E_{J1}\approx D_N\Delta/4.$
As a result, we can use $E_{J1}-E_{J2}\approx D_N V_Z/4$ for the low-transparency limit (in the high-transparency limit it is given by $E_{A1}(\pi)-E_{A2}(\pi)\approx V_Z$, which is also linear in $V_Z$). Then, in  leading order the difference between the
amplitudes of the ABSs is given by the Zeeman term $V_Z$. The effective
phase potential takes the form:

\begin{equation}
U(\phi)=-\frac{1}{2}\left(E_{J1}+(-1)^{n}E_{J2}\right)\cos\phi-\frac{I\phi}{2e},
\end{equation}
which corresponds to Eq. (2) of the main text with the $4\pi$-periodic term set to zero (no MBSs). Here $n=0,\,1$ denotes the occupation of the energetically lower lying ABS;  we assume the
occupation of the higher ABS to be constant (it sticks to the continuum
edge in some vicinity of $\phi=2\pi k$). Then, for $n=0$, the potential
has deep enough local minima to keep the phase localized, as we assume
small currents $I\ll e\left(E_{J1}+E_{J2}\right)\approx eD_N\Delta/2$.
However, the fist excited state $n=1$ may not have local minima,
depending on the ratio $I/(eV_ZD_N)$. At low fields the system is in the running state (no local minima). The effective
critical current depends on the amplitude of the phase potential: $I_{c}=e\left(E_{J1}-E_{J2}\right)\approx eD_N V_Z/4$.
For the running state in the RSJ model the phase evolution is given by
the classical equation of motion:

\begin{equation}
\frac{d\phi}{dt}=2eI_{c}R\left(\frac{I}{I_{c}}-\sin\phi\right)
\end{equation}
\begin{figure}[h]
\includegraphics[width=8cm]{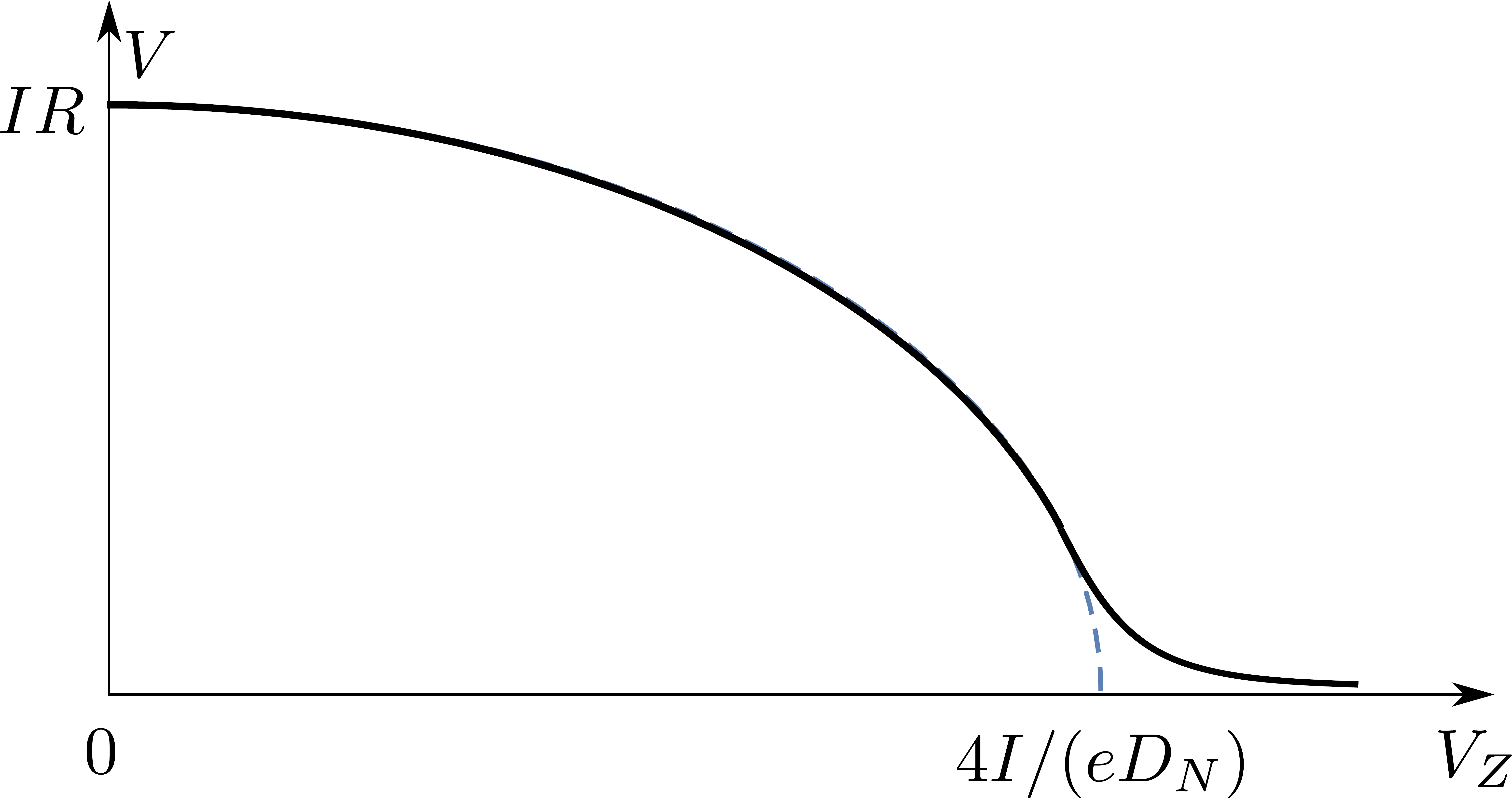}
\caption{A schematic of the average voltage dependence on the applied magnetic field (Zeeman field $V_Z$) in the running state (excitation due to absorption of one quasiparticle). The blue dashed line is given by Eq.~(\ref{eq:volttriv}). The deviation from the dashed line is caused by thermal fluctuations, such as thermally activated phase slips, when the local minima of the phase potential are shallow.}\label{fig:voltage}
\end{figure} \\
The known solution for the average voltage in the running state is~\cite{Tinkham1996}
\begin{equation}\label{eq:volttriv}
\left\langle V\right\rangle =R\sqrt{I^{2}-I_{c}^{2}}=R\sqrt{I^{2}-e^{2}V_Z^{2}D_N^{2}/16}\approx IR-\frac{e^{2}D_N^{2}V_Z^{2}}{32I}R,
\end{equation}
which correspond to Eq.~(3) in the main text for the case of $I \gg I_c$. One should note that now it is the average voltage in the running state, as it is varying in time, which can be easily seen from a simple model of a particle sliding down a one-dimensional washboard potential (without local minima) with friction. The higher the field is, the longer the system spends in the lower voltage regime (i.e the flatter parts of the washboard potential) with short intervals of voltages exceeding the average value (steeper parts of the phase potential). Moreover, we expect a smearing   around $V_Z=4I/(eD_N)$ in Eq.~(\ref{eq:volttriv}), where the voltage goes to zero according to the formula, due to thermal fluctuations (thermally activated phase slips for shallow local minima of the phase potential), see Fig~\ref{fig:voltage}.

As a result, in the trivial state with low magnetic fields quasiparticle poisoning would result in rare switching to the running state with a average voltage given by Eq.~(\ref{eq:volttriv}) and lasting for a short time $\tau_r=1/\Gamma_r\ll\tau_{qp}^{triv}$, where $\tau_{qp}^{triv}=1/\Gamma_t$ is the characteristic time between quasiparticle poisoning events (quasiparticles getting trapped in the lower ABS). However, for larger magnetic fields, $V_Z\gtrsim 4I/(eD_N)$, no voltage will develop in the trivial phase.
\bibliographystyle{apsrev4-1}
\bibliography{QP}